\documentclass{aa}

\usepackage[dvips]{graphicx}
\usepackage{graphicx}
\usepackage{txfonts}
\usepackage{natbib}
\bibpunct{(}{)}{;}{a}{}{,}

\def\modif{}

\begin{document}

\title{Interpreting and predicting the yield of transit surveys: \\
 Giant planets in the OGLE fields}

\author{F. Fressin\inst{1}, T. Guillot\inst{2}, V.Morello\inst{2} \and F. Pont\inst{3}}

\titlerunning{The yield of planetary transit surveys}

\authorrunning{F. Fressin \it et al.}


\institute{Observatoire de la C\^ote d'Azur, Laboratoire Gemini, CNRS UMR 6203, B.P. 4229, 06304 Nice Cedex 4, France\\
           \email{fressin@obs-nice.fr}
	   \and
Observatoire de la C\^ote d'Azur, Laboratoire Cassiop\'ee, CNRS UMR 6202, B.P. 4229, 06304 Nice Cedex 4, France
           \and
           Geneva University Observatory, Switzerland}

\date{A\&A, in press. Received: January 18, 2007, Accepted: April 8, 2007.}

\abstract{
Transiting extrasolar planets are now discovered jointly by
photometric surveys and by radial velocimetry, allowing measurements
of their radius and mass.
We want to determine whether the different data sets are compatible
  between themselves and with models of the evolution of extrasolar
  planets. We further want to determine whether to expect a population
  of dense Jupiter-mass planets to be detected by future more sensitive
  transit surveys.
We simulate directly a population of stars corresponding to the OGLE
  transit survey {\modif and assign them planetary companions based on a list of 153
extrasolar planets discovered by radial velocimetry}. We use a model
  of the evolution and structure of giant planets that assumes that
  they are made of hydrogen and helium and of a variable fraction of
  heavy elements (between 0 and 100 {\modif{$M_{\oplus}$}}). {\modif
  The output list of detectable planets of the
simulations} is compared to the real detections.
We confirm that the radial velocimetry and photometric survey data
  sets are compatible within the statistical errors, assuming that
  planets with periods {\modif between 1 and 2} days are approximately {\modif 5} times less
  frequent than planets with periods {\modif between 2 and 5 days}. We show
  that evolution models fitting present observational constraints 
  {\modif predict a lack of small giant planets with large masses}.
  As a side result of the study, we identify two distinct populations of
  planets: {\modif those with short periods ($P<10d$), 
  which are only found in orbit around metal-rich stars with $\rm{[Fe/H]}>\sim-0.07$, and 
  those on longer orbits ($P>10d$), 
for which the metallicity bias is less marked.} We further confirm the relative absence of
  low-mass giant planets at small orbital distances.
{\modif Testing these results and the underlying
planetary evolution models requires the detection of a statistically significant
number of transiting planets, which should be provided over the next few years by
continued ground-based} photometric
  surveys, the space missions CoRoT and Kepler, and combined radial
  velocity measurements. 
}

\maketitle

\section{Introduction}

Extrasolar planets are now routinely discovered orbiting solar-type
stars by radial velocimetry, but the discovery of {\it transiting}
planets by photometric surveys is just beginning. Although still
marginal, the late success of transit surveys has given an additional
impulse to exoplanetology with the possibility to estimate {\modif the radius,
density and hence composition of extrasolar planets.}

Quantitatively, we know to date 206 extrasolar planets with masses
below 13\,M$_{\rm Jup}$
\citep[e.g.][]{Udry_2007,Butler_2006}. Among those, a list of
14 currently known transiting planets is presented in
table~\ref{table:planets in transit}. These planets have been
discovered by radial velocimetry followed by photometry for 3 of them,
and by photometric surveys for the remaining 11. 

When considering the score of projects devoted to the detection of
planets by transit photometry, the present harvest appears meager. The
discrepancy between predictions \citep[e.g.][]{Horne_2001} and reality
has been attributed to various factors such as: imperfect duty cycle,
a reduced number of stars for which transiting planets are detectable
\citep{Gould_2006} and the presence of correlated noises that can
greatly limit the {\modif detectability of} small planetary transits
\citep{Pont_2006}. {\modif Several generic studies have been conducted
to understand the yield of different transit surveys. \citet{Pepper_2005} 
studied the optimization of transit searches as a function of the 
observational setup, the site properties and the planet properties.
\citet{Gillon_2005} analyzed and compared deep field surveys, considering individual stellar ranges
and observation windows, but did not include the effects of stellar crowding nor time-correlated noises.

\citet{Gould_2006} studied the yield of OGLE survey \citep{Udalski_2002}, the most successful so far in term of number of transiting planets discovered, with a model populating the line of sight with stars drawn from the Hipparcos Catalogue. They estimated with that model the proportion of stars with sensitivity to close-in giant planets to derive from OGLE results the frequency of planets as a function of their period.}
They find that the yield of the OGLE survey is globally consistent with the
detections by radial velocimetry and with planet radii distributed
between 1 and 1.25 jovian radii. The aim of the present work is to
further test these data sets and the underlying physical model by a
forward calculation of transit events with realistic stellar and
planetary populations. In particular, we include up-to-date models of
the evolution and structure of Pegasids (close-in extrasolar
planets) based on models reproducing the observational constraints
from known transiting planets
\citep{Guillot_2005,Guillot_2006}. As a consequence, we should be
able to determine whether the presently known population of transiting
planets represent the ``tip of the iceberg'', i.e. that many more
small, dense extrasolar giant planets exist and await discovery by the
transit method, or whether it is relatively representative of the
global population. 

We first describe the model that is used to simulate transit surveys
in general. In Section~3, we describe more particularly the OGLE
surveys and the hypothesis chosen for their modelling. We then discuss
the results of the simulation. A summary of the main conclusions and
predictions for future transit surveys are provided in Section~5.

\begin{table*}[htbp]
\caption{Known transiting planets by 2006$^\star$}
\centering
\label{table:planets in transit}
\centering
\begin{tabular}{cccccccccc}
\hline\hline
\# & Name &	$M_{planet}$	& $R_{planet}$	& Period	&	a & $M_{\star}$ & $R_{\star}$ & Teff & Metallicity\\
 & & [$\rm M_{Jup}$] & {[$\rm R_{Jup}$]} & [day] & [AU] & [$M_\odot$] & [$R_\odot$] & [K] & [Fe/H]\\
\hline\hline
\multicolumn{10}{c}{OGLE planets} \\
\hline
6 & OGLE-TR-10 &	0.63$\pm 0.14$ &	$1.26_{-0.07}^{+0.07}$ &	3.10129	&	0.04162	&	1.18 $\pm 0.04$	& 1.16$\pm 0.06$	& 6075$\pm 86$	&	0.28$\pm 0.10$  \\   
2&OGLE-TR-56 &	1.17$\pm 0.04$	& $1.32_{-0.06}^{+0.06}$ &	1.211909 & 0.0225 & 1.17$\pm 0.04$	& 1.32$\pm 0.06$	& 6119$\pm 62$	&	0.19$\pm 0.07$\\
5&OGLE-TR-111	& 0.52$\pm 0.13$	& $1.067_{-0.054}^{+0.054}$	& 4.0144479	& 0.047 & 0.81$\pm 0.02$	& 0.831$\pm 0.031$	& 5044$\pm 83$	&	0.19$\pm 0.07$\\ 
3&OGLE-TR-113	& 1.35$\pm 0.22$	& $1.09_{-0.03}^{+0.03}$ & 1.4324757 & 0.0229 & 0.78$\pm 0.02 $	& 0.77$\pm 0.02 $	& 4804$\pm 106$ & 0.15$\pm 0.10$\\
4&OGLE-TR-132	& 1.14$\pm 0.12$	& $1.18_{-0.07}^{+0.07}$ & 1.689868 &	0.0299 & 1.26$\pm 0.03	$ & 1.34$\pm 0.08$	& 6210$\pm 59$	&	0.37$\pm 0.07$\\
\hline
\multicolumn{10}{c}{Other transit survey planets} \\
\hline
7&TrES-1 & 0.76$\pm 0.05$ & $1.081_{-0.029}^{+0.029}$	& 3.0300737	& 0.0393 & 0.89$\pm 0.035$ &	0.811$\pm 0.020$	& 5250$\pm 75$	&	-0.02$\pm 0.06$\\
11&TrES-2 & 1.28$\pm 0.07$ & $1.24_{-0.06}^{+0.09}$	&	2.47063	&	0.0367 & 1.08$\pm 0.08$ & 1.00$\pm 0.05$	& 5960$\pm 100$ & 0.15$\pm 0.03$\\
10&XO-1 &	0.90$\pm 0.07$	& $1.184_{-0.018}^{+0.028}$	& 3.941634 &	0.0488 & 	1.0$\pm 0.03$ &	0.928$\pm 0.033$ &	5750$\pm 13$	&	0.015$\pm 0.03$\\
12&HAT-P-1	&	0.53$\pm 0.04$	& $1.36_{-0.09}^{+0.011}$ & 4.46529 &	0.0551 & 	1.12$\pm 0.09$	& 1.15$\pm 0.09$	& 5975$\pm 45$	&	0.13$\pm 0.02$\\
13&WASP-1 &	0.867$\pm 0.073$	& $1.443_{-0.039}^{+0.039}$	& 2.519961 &	0.0382 & 	1.15$\pm 0.09$	& 1.453$\pm 0.032$	& 6200$\pm 200$ & \\	
14&WASP-2 & 0.88$\pm 0.07$ &	$1.038_{-0.05}^{+0.05}$	& 2.152226 &	0.0307 & 	0.79$\pm 0.08$	& 0.813$\pm 0.032$	& 5200$\pm 200$ & \\	
\hline
\multicolumn{10}{c}{Transiting planets discovered with Radial velocities} \\
\hline
9& HD189733 &	1.15$\pm 0.04$	& $1.154_{-0.032}^{+0.032}$	& 2.218573 &	0.0313 & 0.82$\pm 0.03$	& 0.758$\pm 0.016$	& 5050$\pm 50$	&	-0.03$\pm 0.04$\\
8&HD149026 &	0.330$\pm 0.02$ & $0.726_{-0.064}^{+0.064}$ &	2.87598 &	0.042 &	1.3$\pm 0.1$	&	1.45$\pm 0.1$ &	6147$\pm 50$	&	0.36$\pm 0.05$\\
1&HD209458 &	0.657$\pm 0.006$	& $1.320_{-0.025}^{+0.025}$	& 3.52474859 &	0.047 &	1.09$\pm 0.09$	& 1.148$\pm 0.002$	& 6117$\pm 26$	&	0$\pm 0.02$\\
\hline\hline
\multicolumn{10}{l}{$\rm M_{Jup}=1.8986112\times 10^{30}\,$g is the
  mass of Jupiter. $\rm R_{Jup}=71,492\,$km is Jupiter's equatorial radius.}\\
\multicolumn{10}{l}{OGLE-TR-10: \citet{Bouchy_2005,Udalski_2002,Konacki_2005,Santos_2006,Pont_2006_2} } \\
\multicolumn{10}{l}{OGLE-TR-56: \citet{Konacki_2003,Udalski_2002,Torres_2003}} \\
\multicolumn{10}{l}{\hphantom{OGLE-TR-56: }\citet{Bouchy_2005,Santos_2006,Pont_2006_2}} \\
\multicolumn{10}{l}{OGLE-TR-111: \citet{Pont_2004,Santos_2006,Udalski_2002,Winn_2007,Bouchy_2005}} \\
\multicolumn{10}{l}{OGLE-TR-113: \citet{Bouchy_2004,Udalski_2002,Konacki_2004,Gillon_2006}} \\
\multicolumn{10}{l}{OGLE-TR-132: \citet{Bouchy_2004,Udalski_2003,Moutou_2004_2,Gillon_2007}} \\
\multicolumn{10}{l}{TRES-1: \citet{Alonso_2004,Laughlin_2005,Winn_2007}} \\
\multicolumn{10}{l}{TRES-2: \citet{O_Donovan_2006}} \\
\multicolumn{10}{l}{XO-1: \citet{McCullough_2006,Holman_2006,Wilson_2006}} \\
\multicolumn{10}{l}{HAT-P-1: \citet{Bakos_2006}} \\
\multicolumn{10}{l}{WASP-1: \citet{Collier_Cameron_2006,Shporer_2007,Charbonneau_2006}} \\
\multicolumn{10}{l}{WASP-2: \citet{Collier_Cameron_2006,Charbonneau_2006}} \\
\multicolumn{10}{l}{HD-189733: \citet{Bouchy_2005,Bakos_2006}} \\
\multicolumn{10}{l}{HD-149026: \citet{Sato_2005,Charbonneau_2006}} \\
\multicolumn{10}{l}{HD209458: \citet{Brown_2001,Cody_2002,Wittenmyer_2005,Winn_2005,Knutson_2007}} \smallskip \\
\multicolumn{10}{l}{\# is the label of planets in figures ; they are ranked in the order of their discovery.} \\

\end{tabular}
\end{table*}

\section{Simulating transit surveys}

\subsection{General remarks}

The search for planets in transit in front of their star naturally
arised with the discovery that a non-negligeable fraction of planets orbit
very close to their stars. If orbital planes are randomly
oriented, the probability that a planet will transit in
front of its star at each orbital revolution is:
\begin {equation}
{\cal P}_{\rm transit} \simeq R_\star/a_{\rm planet},
\end {equation}
where $R_\star$ is the stellar radius, and $a_{\rm planet}$ the
planet's orbital semi-major axis. For systems such as 51 Peg b, this
probability is close to 10\%. Because the probability for a solar-type
star to possess such a Pegasid (i.e. a 51 Peg b-like planet, {\modif planets 
semi-major axis up to 0.1 AU}) 
is about 0.5\% \citep[e.g.][]{Marcy_2005}, 1 in 2000 solar-type star should possess a transiting Pegasid. Using current results from radial
velocity surveys and integrating over all periods, we estimate that
about 1 in 1100 solar-type stars possesses a transiting giant
planet. Of course, depending on the magnitudes and field considered,
giant stars may severly outnumber the dwarfs, so that in a real field,
only one in, say, 3000 stars may harbor a transiting giant planet.

Because of the dependence on $a$, and period distribution, most of
the transit events concerning giant planets occur for orbital periods
between 1 and 5 days. The transits typically last for a
couple of hours, as this quantity is weakly dependant on the orbital
period $P$: 
\begin {equation}
\tau_{\rm transit}=1.82 \left(P\over 1\,{\rm day}\right)^{1/3}
\left(M_\star\over M_\odot\right)^{-1/3}\left(R_\star'\over R_\odot\right)\
\rm hours,
\end {equation}
{\modif where $R_\star'$ is the length of the cord traced on the stellar disk by the planet's trajectory.}
(more precisely: $R_\star'=R_\star \cos b + R_{\rm planet}$,
where $b$ is the impact parameter of the transit). 

The {\modif depth} of the transits themselves is directly given by the
ratio of the planetary to the stellar disk surfaces:
\begin {equation}
{\cal R}_{\rm transit} \simeq (R_{\rm planet}/R_\star)^2.
\end {equation}
This value is of order 1\% for a Jupiter-size planet orbiting a
Sun-like star. For an F-type star with radius $\sim 1.2\rm\,R_\odot$,
the ratio decreases to 0.7\%. Furthermore, transiting giant planets
discovered so far have radii between $0.72$ and $1.44\,R_{\rm Jup}$
(see table~\ref{table:planets in transit}). Allowing for stellar radii
to vary between 0.8 and $1.3\,R_\odot$ (a typical range, in magnitude
limited surveys), this implies that we should expect ${\cal R}_{\rm
transit}$ to vary between 0.3\% and 3\%, for giant planets only.  
The lower limit is in reality even smaller because for
detection purposes we have to account for the fact that planets also
orbit stars that are in multiple systems (like HAT-P-1), and hence a
dilution factor may {\modif apply}. {\modif Although grazing transits 
are ignored in this simple analysis, they are included afterwards in our simulations.}

This altogether implies that in order to detect transiting giant
planets, many thousands of dwarf stars have to be monitored over
periods of weeks for a photometric precision reaching {\modif below} a fraction of a
percent on an equivalent integration time of about one hour. This is
typically done by following a relatively dense stellar field over a
long time with a stable telescope, and using a camera equiped with a
good CCD camera.

\subsection{Principle of the simulations}

On paper, the simulation of the forward problem is simple: one has to
generate a complete stellar field, or obtain it from observations,
put it on a discrete grid (the CCD), include on probabilistic
arguments the planetary companions, calculate lightcurves including
the various sources of noise, and determine which events are
detectable. This is the principle of CoRoTlux, a code {\modif we} first developed
to predict the transit yield of CoRoT space telescope \citep{Baglin_2002} 
and quantify the need for follow-up observations, which is here applied to the case of OGLE. 

The interesting point of such a forward simulation is the possibility
to include relatively easily fine details such as the fact that
planets are found more frequently around metal-rich stars, or,
on the basis of planetary evolution models, the fact that young planets
orbiting close to bright stars will be larger than old planets
orbiting smaller stars at larger orbital distances. This requires
however that a relatively large number of physically relevant
parameters (for example, the mass, size, metallicity, age of the
stars) be properly defined. 

We further detail the assumptions that we made for these simulations
by describing how we generate the stellar and planetary populations,
and how we attempt to include realistic sources of noise.

\subsection{The stellar population}

\subsubsection{Main targets and background stars}

Stellar fields differ enormously in terms of densities and stellar
populations. It is therefore most important to properly account for
this in order to simulate any given transit survey.

It would be very appealing to use direct observations as much as
possible to closely match the observed target fields.  But as we will
see hereafter, many different characteristics of the stars (stellar
metallicity, age and subtype ...) are required, and these are
difficult to obtain with generic observations. We therefore adopt
the following procedure:
\begin{itemize}
\item The observed stellar densities are obtained from stellar counts
  by magnitude, on the real stellar fields {\modif (see \S~\ref{sec:OGLE survey})}
\item The characteristics of the stars are obtained following a
  Monte-Carlo method using the output of the Besan\c{c}on model of the
galaxy \citep{Robin_2003} obtained for the proper location of the survey.
\item Where stellar counts are not available, or uncomplete (i.e. for
  faint stars), we use both stellar counts and characteristics from
  the Besan\c{c}on model.
\end{itemize}

Specifically, we keep track of the following parameters obtained directtly from
the Besan\c{c}on model: 
\begin{itemize}
\item The mass of each star, used to compute orbital parameters of the
  transiting object;
\item The apparent magnitude of the star in the observed spectral
  range (the I filter in the case of the OGLE survey);
\item The V magnitude of the star, important to qualify the
  confirmability of a transit event with radial velocimetry;
\item The surface temperature of the star
\item The luminosity of the star, calculated from its absolute magnitude;
\item The radius of the star, calculated from total luminosity and
  effective temperature.
\end{itemize}

\begin{figure}
\centering
\includegraphics[width=0.8\linewidth,angle=0]{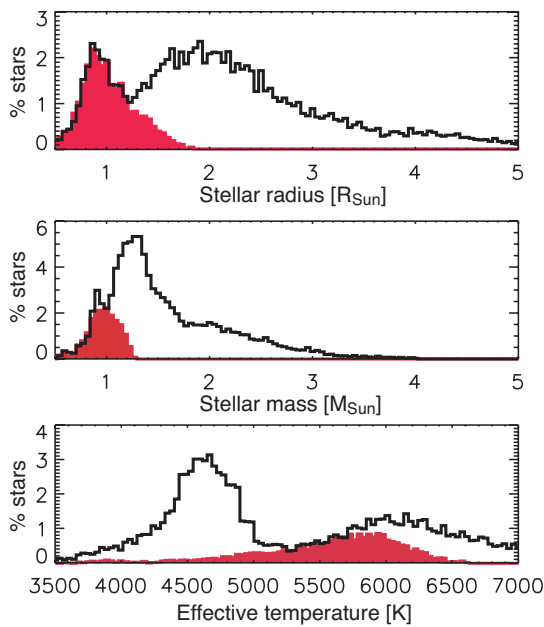}
\vspace*{0.5cm}
\caption{From top to bottom: Distribution functions for the radii,
   masses and effective temperatures for our fiducial stellar
   population corresponding to the simulated OGLE Carina field. The
   black line represents the ensemble of stars in the field. The filled
   red region is a subset for dwarf stars with stellar type F4 and
   later, as these are the only stars for which a transiting planet has
   a reasonnable chance of being detected by present-day transit surveys.  }
\label{fig:stars_r_m_t}
\end{figure}

The mass, and effective temperature of the stars are distributed linearly 
around values given by the Besan\c{c}on model (at $\pm 20 \%$). Figure~\ref{fig:stars_r_m_t} 
shows a simulated distribution of stars for the OGLE Carina field. 
The ensemble of dwarf stars with
types F4 and later are highlighted as these represent targets for
which planetary transit events are detectable, and, within
observational limits, confirmable by radial velocimetry.

The metallicity distribution is obtained from the model of \citet
{Nordstrom_2004}, which is based on the Geneva-Copenhagen survey of
the Solar neighbourhood. These authors find that the distribution of
the metallicities [Fe/H] is well approximated by a Gaussian function
with a mean of $-0.14$ and a dispersion of $0.19$ dex. We use this
gaussian distribution and choose to ignore possible dependencies
between stellar parameters (e.g. masses, ages...) and the
metallicities. {\modif (The link between stellar type and metallicity 
appears to be negligible for F4 and later types anyway (F. Th\'evenin,
pers. communication 2007)).}

\subsubsection{Binary and triple systems}

Multiple stellar systems are important especially because of the
possibility that stellar eclispes mimic planetary transits \citep{Brown_2003}. 
However, we choose to defer this problem to a later
article. Multiple systems are taken into account anyway because they
can yield a dilution of the planetary transit events that makes them
more difficult to detect. Planets may be present indifferently on the
primary, secondary or {\modif tertiary} components of a stellar system. (However,
we find that only planets around the primary targets have a
non-negligible chance of being discovered by current ground based
photometric survey.) 

Specifically, following \citet{Duquennoy_Mayor_1991}, we consider
that $50\%$ of the stars are binaries and $20\%$ of those are
ternaries. {\modif Multiple systems are considered as individual stars at the same position on the CCD}. 
We choose to estimate their properties more simply
than for the other stars, on the basis of DM91:
\begin{itemize}
\item {\modif We randomly add companions to the initial draw of primary stars, 
without changing their properties. The total mass and luminosity of each multiple system
is thus slightly higher than initially.}
\item The mass ratio (primary/secondary) is defined as a gaussian of
  median value 0.23 and a full width at half maximum of 0.42. Outside
  a range of 0.05 and 1, we redraw the mass ratio. 
\item The radius is defined as $R_\star=R_\odot (M_\star/M_\odot)$ when
  $M_\star \le M_\odot$ and $R_\star=R_\odot (M_\star/M_\odot)^{1/2}$
  otherwise. 
\item The luminosity is assumed to be proportional to $M^2$ so that:
  $l_{\rm secondary}=l_{\rm primary}(M_{\rm secondary}/M_{\rm
  primary})^2$. 
\item Other stellar parameters are calculated on the basis of these
  ones and of those of the primary component (same age, same distance,
  same metallicity).
\item Triple components are treated with the same method as binaries, 
\modif{and are defined relatively to the primary star.} 
\end{itemize}

\subsection{The planetary companions}
\label{sec:planets}

With more than 200 planets known to orbit stars other than our Sun, we
are beginning to have a rather precise view of at least part of this
population. We can expect that biases on the detections are small in
the case of massive planets (the mass of Saturn and more) and planets 
that are relatively close to their star (orbital
 distances smaller than $\sim 1\,$AU). These two conditions happen to 
match quite exactly the requirement for
detectability by transit photometry, with one assumption: that only
massive giant planets can have large radii. Although not proven, this 
assumption seems quite reasonnable. 

Hence we choose to focus this study on this well-characterized
population of objects. From the current list of $209$ detected
exoplanets, we select the ones discovered by radial velocities {\modif with
mass higher than $0.3$ Jupiter masses and known host star metallicity}. 
Our list of planets includes $153$ objects, to which we may
add very-close in planets detected by transit photometry, as discussed
below. We are using this list as representative of an unbiased sample
of giant planets known from radial velocimetry, even though planetary 
distribution models may have been made from slighlty different samples.

\subsubsection{Planet incidence}
\label{sec:probability}

\begin{figure}
\centering
\includegraphics[width=0.9\linewidth,angle=0]{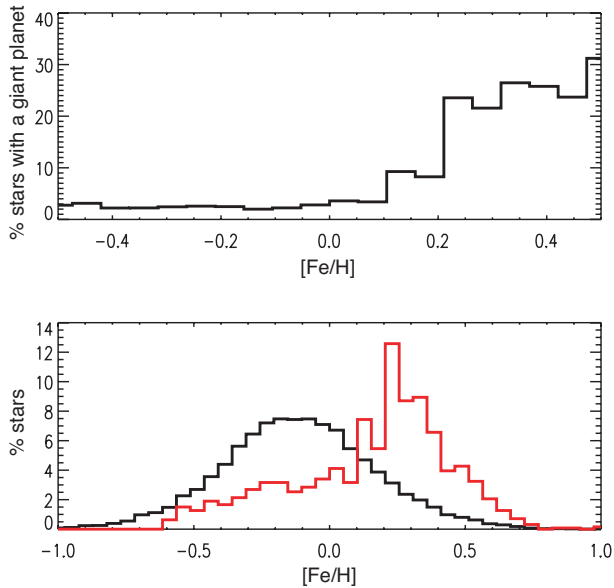}
\caption{Upper panel: Probability for a solar-type star to possess a
  giant planet companion as a function of the stellar metallicity
  \citep[from][]{Santos_2004}. Lower panel: Relative normalised
  distributions of stellar metallicities for stars in the field (black
  line), and for stars with a giant planet companion (red line).}
\label{fig:metallicity_stars}
\end{figure}

A first important step is the determination of the probability for a
star to harbor a planet. As shown by numerous studies
\citep{Gonzalez_1998,Santos_2004,Fischer_valenti_2005}, this
probability depends mostly on the metallicity of the parent
star. Figure~\ref{fig:metallicity_stars} shows one such probability
function, as well as the result in terms of planet counts on a
simulated stellar field.

In this work, we will use the dependency from \citet{Santos_2004}
shown in Fig.~\ref{fig:metallicity_stars}. Several points are to be
considered however:
\begin{enumerate}
\item This probability relation is only valid for solar-like stars,
  i.e. F, G, K dwarf stars. Although there are strong indications that
  it may change for other stars (e.g. M dwarfs), we will assume it to
  hold independently of stellar properties. This assumption is sufficient
  because F, G and K dwarf stars form by far the majority of stars
  with detectable planets in photometric surveys.
\item This relation has been calculated independently of the
  properties of the planetary companion, in particular orbital
  distance. Because in our case we are strongly biased towards
  short-period planets, the distribution may be different. This point
  will be considered in \S~\ref{sec:metallicity}.
\item The possibility of multiple planetary systems is not
  considered. This approach is justified because the probability that
  several planets belonging to the same system are transiting planets
  is small for giant planets.
\end{enumerate}

\subsubsection{Planetary masses and orbits}
\label{sec:masses and orbits}

The masses and orbital characteristics of the planet population are
inferred almost entirely from the present radial-velocimetry
surveys. This technique yields an accurate determination of the
orbital period, and less accurately, of the eccentricity of the
orbit. It also yields the value of the mass of the planetary
companion times the sine of the orbital inclination from the knowledge
of the mass of the parent star. With these values, we can then derive
from a random inclination of the orbital planes the planets that are
transiting and those that are not as well as the characteristics of
their orbit. 

We test several approaches for the derivations of these quantities:
\begin{itemize}
\item {\it An analytical model:}
In this approach, we consider independantly
the planet period and its mass. The period of the planet $\Pi$ follows
the model of \citet{Brown_2003}, the probability density $\cal P$ from a
piecewise linear fit to the distribution ${\cal P}(\log \Pi) = \{0.509, 0.165,
0.533\}$ for three period intervals bounded by $\log \Pi = \{0.43,
0.65, 2.3, 3.5\}$. The distribution in mass is linear in log from
$0.3$ to $10$ Jupiter masses \citep{Zucker_2001}. There is no
dependency of these two parameters linked to metallicity. 

\item{\it The radial velocity mass-period ``carbon-copy'' model:}
A second approach is to make direct use of the list of planets
discovered by radial velocimetry. This is possible because in terms of
masses and orbital periods the list is almost unbiased for the objects
that we consider (massive enough to be above detection thresholds, and
with periods much shorter than the lifetimes of the surveys). In this
case, we select planets randomly in the RV list, and then allow for a
small random deviation of their mass and period (a uniform deviation
from $-20\%$ to $+20\%$) in order to avoid too much clustering on the
same value. This is particularly important in the case of the period
because of the importance of stroboscopic effects in planetary
transits \citep[e.g.][]{Pont_2005}.

\item{\it The radial velocity mass-period-metallicity ``carbon-copy''
  model:}
As a modification to the previous approach, we also consider using the
metallicity entry in the RV list, because of correlations between
metallicity and orbital period that are otherwise not taken into
account (see discussion in section~\ref{sec:results}). We proceed slightly
  differently however than for the mass and orbital period because of
  the limitations caused by the finite number of planets in the RV
  list. In this case, we choose to split the list into two
  low- and high-metallicity lists, and then select the mass and
  periods in the relevant list. Our fiducial cutoff value is
  [Fe/H]=$-$0.07. 
\end{itemize}

\begin{figure}
\centering
\includegraphics[width=0.8\linewidth,angle=0]{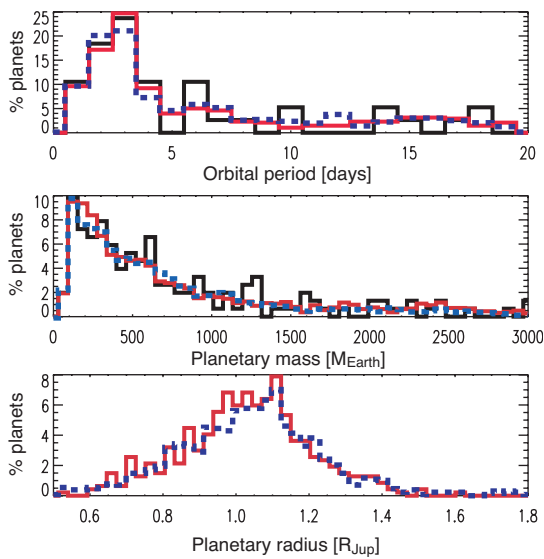}
\vspace*{0.5cm}
\caption{From top to bottom, distributions of orbital periods, masses
  and radii, respectively, of the planets observed by radial
  velocimetry (black lines), simulated as part of the mass-period
  ``carbon copy'' model (red lines), and simulated as part of the
  analytical model (dotted blue lines) (see text). }
\label{fig:planets_p_m_r}
\end{figure}

Figure~\ref{fig:planets_p_m_r} shows a comparison between
observations, the carbon copy model and the analytical model. It is
interesting to notice at this point that the carbon copy and
analytical models are essentially indistinguishable in these
diagrams. The differences with the observations arise only because of
our choice to smear the masses and orbital periods when generating our
planet population. 

Last but not least, we have to consider the existence of planets that
orbit extremely close to their star, with periods shorter than 2 days,
as discovered by transit surveys (see table~\ref{table:planets in
  transit}). Companions with such short orbital periods have been
discovered by radial velocimetry in two occasions: {\modif HD~41004~b, and Gliese~876~d}, 
with respective masses $18.4$ and $0.023$
Jupiter masses. These objects are outside the mass range considered
for this study, and therefore, there is no giant planets with periods
shorter than 2 days in the present radial velocimetry list. In order
to account for these very close-in planets anyway, we add the planets with 
periods smaller than 2 days
discovered by transit photometry to the list, but with a small
tunable probability weight. The fiducial value of this parameter is
set so that, on average, the planet list contains one and a half such planet
(added to the list of 153 RV planets described in \S~\ref{sec:planets}). 
Tests on the effect of this parameter are presented in \S~\ref{sec:statistics}.

Our fiducial model is the mass-period-metallicity carbon copy
model, includes addition of very-close in planets {\modif and it is that model which is used 
in all cases except where specified otherwise}. Other approaches are also
tested depending on the model to highlight particular points. 

\subsubsection{Physical characteristics and the planetary evolution  
model}
\label{sec:irradiated}

Because we are focussing on planets more massive than Saturn, we
expect most of them to be made of a significant amount of hydrogen and
helium. These giant planets thus undergo a progressive contraction and
cooling that depends on four quantities: their age, mass, the amount
of flux the planet receives from the central star, and the global
amount of heavy elements in the planet \citep[e.g.][]{Guillot_2005}.

Models attempting to reproduce the radii of known transiting giant
planets have however had problems in explaining the large radii of
some of them \citep{Bodenheimer_2001,Guillot_Showman_2002,Baraffe_2005,Laughlin_2005}. 
Several possibilities have been
proposed to explain the discrepancy. We can separate them into two  
categories:
\begin{itemize}
\item Mecanisms invoking chance configurations of the planetary orbits
   in the case of these anomalously large planets: this includes
   the tidal circularization of an eccentric orbit \citep{Bodenheimer_2001}, 
   and tidal dissipation for a planet locked in a Cassini
   spin-orbit resonnance with the star \citep{Winn_Holman_2005}.
\item Effects that would apply to all planets, including problems with
   the equations of state or opacities, and the dissipation by stellar
   tides of kinetic energy first generated in the atmosphere \citep{Showman_Guillot_2002}.
\end{itemize}

The first mecanisms appear to have a low probability of occurence
\citep{Laughlin_2005,Deming_2005,Levrard_2007}. The second possibility therefore 
seems more likely, but requires in some case the presence of
relatively large masses of heavy elements to reproduce the observed
radii.

\begin{figure}
\centerline{\resizebox{10cm}{!}{\includegraphics{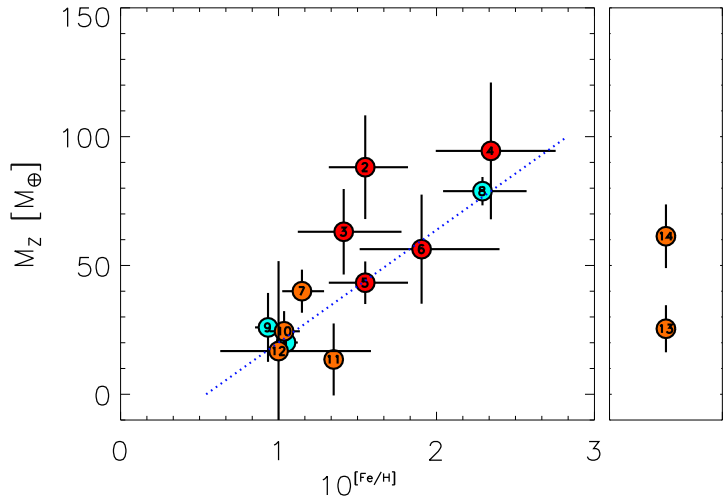}}}
\caption{Mass of heavy elements in transiting Pegasids known by 2006
   as a function of the metal content of the parent star relative to
   the Sun. The mass of heavy elements required to fit the measured
   radii is calculated on the basis of evolution models including an
   additional heat source slowing the cooling of the planet. This heat
   source is assumed equal to $0.5\%$ of the incoming stellar heat
   flux \citep{Showman_Guillot_2002}. Horizontal error bars correspond
   to the $1\sigma$ errors on the [Fe/H] determination. Vertical error
   bars are a consequence of the uncertainties on the measured
   planetary radii and ages. {\modif The metallicity of recently discovered planets
   WASP-1 and WASP-2 (right panel) is not precisely known.} 
   The dotted line corresponds to a best fit model. [Adapted from \citet{Guillot_2006}].}
\label{fig:correlation}
\end{figure}

A model-dependant estimate of the masses of heavy elements present in
the currently known transiting Pegasids is shown in
Fig.~\ref{fig:correlation}. This model relies on the hypothesis that
0.5\% of the absorbed stellar flux is used to generate kinetic energy
that is subsequently dissipated deep into the planetary interior \citep{Guillot_Showman_2002}. 
As proposed by \citet{Guillot_2006}, there
appears to be a correlation between the amount of heavy elements
present in the planet and the metallicity of their parent star.

This correlation has to be ascertained, but we choose for our fiducial
model to adopt a unique relation between metallicity and mass of heavy
elements (treated as a central core in our calculations),
corresponding to the dotted line in Fig.~\ref{fig:correlation}:
\begin{equation}
M_Z=43.75\times 10^{\rm [Fe/H]}-23.7\ \rm M_\oplus.
\end{equation}
We limit the range of possible values of $M_Z$ to $[0,100M_\oplus]$.

Similarly, we adopt a simple boundary condition for our evolution
calculations:
\begin{equation}
T_{\rm 1 bar}=1.25 T_{\rm eq0},
\end{equation}
where $T_{\rm 1 bar}$ is the temperature at the 1 bar pressure level
and $T_{\rm eq0}$ is the equilibrium temperature for a zero albedo
(see Guillot 2005 for a description), calculated as a function of
stellar effective temperature and radius and planet semi-major axis.

For simplicity, and because it yields only minor changes on the
results, we further choose to neglect the time-dependence in the
evolution calculations, and to adopt the equilibrium radius, or the
{\modif 10\,Gyr} solution, whichever occurs first.

Practically, planetary radii are obtained from interpolations in a {\modif
table based on three parameters: the planetary mass ranging 
from $100$ to $3000 M_\oplus$, the core mass from $0$ to $100 M_\oplus$ and the 
equilibrium temperature from $100$ to $2000$ K.
Models were not calculated beyond these values of $T_{\rm eq}$ because of convergence
problems. However we allowed for a slight extrapolation of the tables to $2600$ K to account for
rare extremely hot planets. \footnote{An electronic version of the table is available at
www.obs-nice.fr/guillot/pegasids/} }

Similarly, because of convergence problems for planets with {\modif small 
total masses and large core masses}, we limited the mass of the core to
75\,M$_\oplus$ for planets with masses smaller than
$180$\,M$_\oplus$. More detailed work is required to better simulate
this parameter space, including planets less massive than considered
in this study.

\begin{figure}[htbp]
\centering
\includegraphics[angle=0,width=0.9\columnwidth]{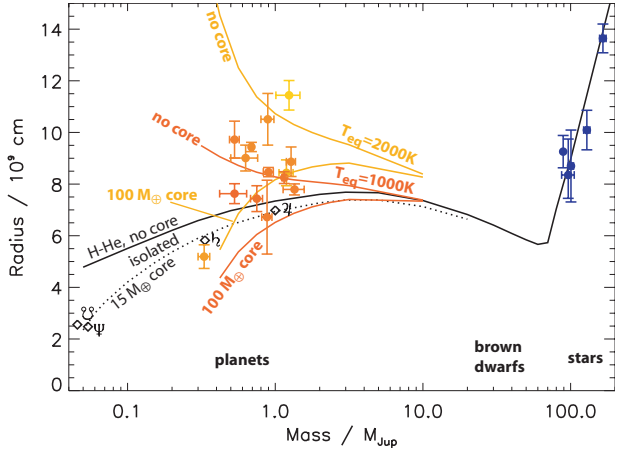}
\caption{Theoretical and observed mass-radius relations. The black
   line is applicable to the evolution of solar composition planets,
   brown dwarfs and stars, when isolated or nearly isolated (as
   Jupiter, Saturn, Uranus and Neptune, defined by diamonds and their
   respective symbols), after 5 Ga of evolution. The dotted line shows
   the effect of a $15 M_\oplus$ core on the mass-radius relation. Orange
   and yellow curves represent the mass-radius relations for heavily
   irradiated planets with equilibrium temperatures of 1000 and
   2000\,K, respectively, and assuming that 0.5\% of the incoming
   stellar luminosity is dissipated at the center (see
   section~\ref{sec:irradiated}). For each irradiation level, two cases
   are considered: a solar-composition planet with no core (top curve),
   and one with a $100 M_\oplus$ central core (bottom curve). The transiting
   extrasolar giant planets for which a mass and a radius was measured
   are shown with points that are color-coded in function of the
   planet's equilibrium temperature. The masses and radii of very low
   mass stars are also indicated as blue points with error bars.}
\label{fig:mass_rad}
\end{figure}

Figure~\ref{fig:mass_rad} shows examples of radii obtained for
$T_{\rm eq}=1000$ and $2000$, K, and core masses of $0$ and $100 M_{\\oplus}$,
respectively, compared to available measurements.

\subsection{Modeling transit events and their detectability}

We now descibe how this population of stars, planets and their
interactions during transits are modelled to reproduce real
observations.

\subsubsection{PSFs and CCDs}

Each image of a star is spread by the atmosphere and by the telescope
to grow to a specific size and shape when reaching the CCD in the
focal plane of the telescope, the so-called point spread function
(PSF). The CCD being composed of many discrete pixels, these PSFs are
then effectively discretized, so that the signal to be analyzed for
any given star is composed of many different lightcurves corresponding
to the many pixels over the size of its PSF. A further complication
arises from the fact that the stellar fields generally chosen by
transit surveys are dense, so that many PSFs overlap. Recovering
individual stellar light curves from real data is a complex
problem. Two popular methods are aperture photometry
\citep{Stetson_1987} and image subtraction \citep{Alard_1998}. (An adaptation
of the latter was used to extract the OGLE lightcurves). 

A refined simulation could include possible spatial and temporal
variations of the PSFs, and a realistic data reduction pipeline. In
our case, we choose to simplify the problem by relying on a posteriori
analyses of real light curves to provide us with a global noise
budget. We however include background stars because of the important
effect of signal dilution. 

In order to do so, we assume that the PSFs are gaussian with a uniform,
constant FWHM. (Non-gaussian PSFs are not difficult to include but we
tested in the OGLE case that for a fixed equivalent FWHM, they have a
negligible effect on the resulting signal-to-noise ratio of simulated
transit events). We consider for each target of the survey the global
flux from the main star and the background stars in its neighborhood
up to magnitude $22$ in the spectral band of observation. The neighborhood 
zone for background stars is defined as a circle of
diameter equal to 4 times the PSF's FWHM around the photocenter of
each target star. Each background star whose photocenter is located in
that zone is taken into account for the calculation of the global
flux. The global flux is calculated as the sum of the pixels located at less 
than twice the FWHM of the central star.

We thus simulate aperture photometry when image subtraction was used for OGLE 
\citep{Udalski_2002}. The choice of the reduction algorithm indeed affects the 
sensitivity obtained from real observations. In our simulations, i.e. a 
relatively idealized case, it would have marginal effects since realistic noises
are included a posteriori from the analysis of real lightcurves (see hereafter).



\subsubsection{Noise budget and event detectability}


We choose to separate noise sources into two categories:
\begin{itemize}
\item `White noise' sources, following gaussian and Poisson laws. 
The main source of white noise is the photon noise of target stars and their background. 
The level of white noise for a given target star 
is obtained from the simulation of the flux of that star and its background in the photometric 
aperture. 
\item `Red noise', or systematic effects on photometry, that undergo temporal correlation.
The structure of these systematics in the OGLE photometry have been explored in details 
by \citet{Pont_2006}. These noise sources are both instrumental (jitter and
breathing of the CCD, frequency spectrum of stellar field moves on the camera,
change of the PSF shapes accross the CCD during the night), and
environmental (differential refraction and extinction, changes of
airmass and sky brightness, temperature changes). Rather than trying to simulate instrumental 
and environmental noise sources accurately, which is difficult with the relatively poor
knowledge we have of the time spectrum of their combined effects, we use the
effective global `red noise' measurements of OGLE-III survey real light 
curves {\modif mentionned in \citet{Pont_2006}}, which consider the combined effect of these noise sources. 
\end{itemize}

\citet{Pont_2006} calculated that, in the presence of a  
mixture of white and red noise (i.e. accounting for photometric  
systematics), the detection threshold for a transit survey is well  
described by a limit on the signal-to-noise ratio defined as:
\begin {equation}
S_r^2={d^2 n^2\over \sum_{k=1}^{Ntr}
  nk^2({\sigma_w}^2/n_k+{\sigma_r}^2)}
\label{eq:S_r}
\end {equation}
where $Ntr$ is the number of transits sampled, $n_k$ the number of
data points in the $k$-th transit. $\sigma_w$ and $\sigma_r$ are the
standard deviation of measurement points of white and red noises,
respectively, $d$ is the event depth and $n$ the total number of
measurement points {\modif during the transit}. Specifically, we obtain $n_k$ by counting for each
transit the number of observation points between the middle of ingress
and the middle of egress.

Equation~\ref{eq:S_r} makes the disctinction between ``white'' noise sources  
that decrease with $n^{1/2}$, where $n$ is the number of successive  
measurements, and ``red'' noise sources that are limited by temporal  
correlation. \citet{Pont_2006} indeed show that taking the red noise into  
account makes a large difference on the detection threshold -- in  
general as well as in its dependence to the planet parameters -- and  
that models based on the assumption of white noise can be poor  
approximations of the actual detection threshold.

\section{The OGLE survey: input parameters}

\subsection{Basic parameters and observational procedure}
\label{sec:OGLE survey}

The Optical Gravitational Lensing Experiment (OGLE) has done {\modif 6} observation campaigns
looking for transiting planets towards different fields of view {\modif from 2001} \citep{Udalski_2002}. 
It took place at the Las Campanas Observatory, Chile, using the 1.3 m Warsaw 
telescope and the 8k MOSAIC camera, with a total field of view of $0.34^{{\circ}2}$. 
All observations were made through the I filter. We assume for our PSF simulation 
an average seeing of $1$ arcsec.

{\modif We analyze in this work the first three OGLE-III observation campaigns 
dedicated to transit search, as their treatment, analysis and follow-up (with current data 
processing pipelines) has been completed:

\begin{itemize}
\item OGLE-III-1 (June $12$ to July $28$, $2001$, described in \citet{Udalski_2002, Udalski_2002b}). More than $800$ images of
three fields in the direction of the galactic bulge were collected
within $32$ nights. The exposure time was $120$ s, and each field was
observed every $12$ min.

\item OGLE-III-2 (February $17$ to May $22$, described in \citet{Udalski_2003}). More than $1100$
images of three fields located in the Carina region of the galactic
disk were collected in $76$ nights. The exposure time was $180$ s,
and the temporal resolution was about $15$ min.

\item OGLE-III-3 (February $12$ to March $26$, described in \citet{Udalski_2004}). The photometric data
were collected during $39$ nights spanning the $43$ days of the
survey. Three fields of the galactic disk were observed with a time
resolution of about $15$ min. The exposure time was $180$ s.

\end{itemize}}

In this article, we will refer to these three observation campaigns
respectively as `Bulge', `Carina', and `Centaurus' fields.

The simulations include the real observation windows of each survey,
as kindly provided by A. Udalski. For any transiting planet in the
simulation, the number of effectively observed transits is used in
eq.~\ref{eq:S_r}. 

In order to construct a realistic stellar population, we use the
stellar counts per magnitude range obtained by
\citet{Gould_2006} {\modif based on OGLE-II data, which have calibrated photometry}. 
We then randomly select that number of stars per
magnitude from the Besan\c{c}on model. In order to test the validity
of our approach, we calculated the fraction of ``stars for which
transits are detectable'' {\modif and compared it} to the one determined by
Gould. This fraction is defined for a given magnitude range as the
number of stars around which a planet orbiting edge-on with $r=1.2
R_{jup}$ and $a = 7.94 R_\odot$ can be detected, divided by the total
number of stars of that magnitude. As shown by
table~\ref{table:fraction of stars probed}, there is an excellent
agreement between our results and those of \citet{Gould_2006}. Note
however that for the global simulation, the {\it complete} star list
is used as the above definition for suitable stars is restricted to
planets of a given size and orbital distance.

\begin{table}
\caption{Fraction of stars suitable for transit detection}
\label{table:fraction of stars probed}
\centering
\begin{tabular}{ccccc}
\hline\hline
& \multicolumn2{c}{Carina} & \multicolumn2{c}{Bulge}\\
$V_{max}$ & {Gould 2006} & This work &  {Gould 2006} & This work\\
\hline
15.5 & 0.11 & 0.16 & 0.138 & 0.141\\
16 & 0.14 & 0.16 & 0.125 & 0.128\\
16.5 & 0.16 & 0.15& 0.098 & 0.105\\
17 & 0.16 & 0.15 & 0.068 & 0.080\\
17.5 & 0.16 & 0.14 & 0.041 & 0.052\\
\hline\hline
\end{tabular}
\end{table}

We calculated the average flux for target stars, companions and all the 
background stars near enough to contribute to the target PSF. 
We then checked that the average photon noise simulated for target stars at 
a given magnitude was close to real values obtained in
OGLE light curves at given magnitude presented in figure 4 of \citet{Pont_2006}. 


\subsection{Modelling the detection threshold}
\label{sec:noise levels}

The candidates in the OGLE survey have been identified with the BLS  
transit-search algorith of \citet{Kovacs_2002}. A subset of the  
candidates selected with cuts in the $\alpha$ and SDE parameters of  
the BLS were examined by eye, and only the best were included in the  
final list. Therefore, the selection threshold is mainly defined by  
subjective appreciation from an experienced specialist.
Recently, \citet{Pont_2006} have pointed out that the effective  
detection threshold of ground-based transit surveys such as OGLE is  
importantly affected by correlated noise (photometric systematics). The  
subjective selection of candidates is in large part necessary because  
of the presence of this correlated noise, which produce many spurious  
detections near the threshold.
   \citet{Gould_2006} chose to model the OGLE selection threshold  
with an  $\alpha>12$ cut (alpha is equivalent to the signal-to-noise  
ratio of the transit signal assuming uncorrelated noise and  
homogeneous distribution of the data points in phase). \citet{Pont_2006} 
have included the effect of correlated noise in the signal-to- 
noise calculation and found that the OGLE selection could be  
better described by a threshold of 8 on the signal-to-noise ratio of the  
transit signal calculated including correlated noise ("$S_r$" in  
their notation, see Sec. 2.5.2), and without the assumption of  
homogeneous coverage. While the two thresholds have similar effects  
on the global number of planet detection, they have a very different  
dependence on some parameters, such as planet period and host star  
magnitude. Since the objective or our study is to examine the  
detection statistics in {\modif a multi-dimensional parameter space}, we use the \citet{Pont_2006}  
description of the OGLE detection threshold.

To calculate $S_r$, one needs an assumption on the level of red noise  
present in the photometry. Following \citet{Pont_2006}, we use a single-parameter  
description and assume $\sigma_r$ = 3.6 mmag in the Bulge fields, $\sigma_r$= 3.1 
mmag in the Carina and Centaurus fields, and $\sigma_r$= 2.1 mmag  
in all fields after application of decorrelation algorithms.

\subsection{Confirmability of transit-like events with follow-up}

High-resolution spectra allow the confirmation of the planetary
events if spectral lines are deep enough. Several scenarios make the
follow up of candidates too difficult: early type stars have lines too
weak and too broadened by rotation (type F4 and earlier). Stars with
magnitudes $V>17.5$  
are too faint for present instruments and telescopes.
This is the limit at which observers estimated not being able to
provide low-metallicity stars. {\modif Those stars} having weaker lines, could also be
difficult to follow correctly, but as planets are unlikely to be found
near this kind of stars in our model, we did not take that parameter
into account.

To simulate the feasability of follow-up, we only considered in
CoRoTlux the stars matching the criteria $V < 17.5$
and of type F4 and later.

\section{Results of the simulations}
\label{sec:results}

We present hereafter runs for the three OGLE-III campaigns for the
fields in the {\modif Galactic} bulge, in Carina and in Centaurus. In order to
obtain a statistically significant population of detected planets, the
simulations were run multiple times. 

We first examine the consistency between the models and observations
for relevant physical variables. In doing so, we choose to compare our
model population to the global population of transiting planets
discovered by OGLE and other surveys. {\modif There is a
slight inconsistency in assuming that} the parameter comparison is almost
independant {\modif of} the type of survey and observational strategy. In some
cases, this is not true, and a clear distinction between
the OGLE planets and the other detections has to be made.

We then discuss the problem of the detection statistics, whether
observations and models are consistent, and whether a constraint on
the (low) frequency of very close-in planets can be deduced. 

{\modif
\subsection{Deviation of OGLE planets from maximum likelihood of the simulations}
\label{ss.ML}

We use a Maximum-Likelihood (ML) technique in order to test whether model results and observations 
agree with each other. We do the tests in two-dimension
spaces, in order to qualify possible correlation and exclusion zones. 
The ML technique is our method of choice as it is a powerful tool for
fitting a model to a multi-dimentionnal independant-data distribution \citep{Lyons_1986}.

Instead of determining an approximate analytical law fitting our results, we use the results of a very large Monte-Carlo draw ($1000$ times the whole OGLE survey,
corresponding to $\sim 9000$ planets) to get a map of the density of probability in each 2-dimension
grid. We bin our data on a 20x20 grid as a compromise between resolution of the models and characteristic
variations of the parameters.\footnote{Tests with different grids yield small variations of the results. As an example, the mass-radius deviation from maximum likelihood is respectively 0.67, 0.65 and 0.72 $\sigma$ for 20x20, 30x30 and 40x40 grids.}
The probability of an event in each bin is considered equal to the normalized number of draws in that bin.

\begin{figure*}
\vspace{0.52cm}
\centerline{\resizebox{13cm}{!}{\includegraphics{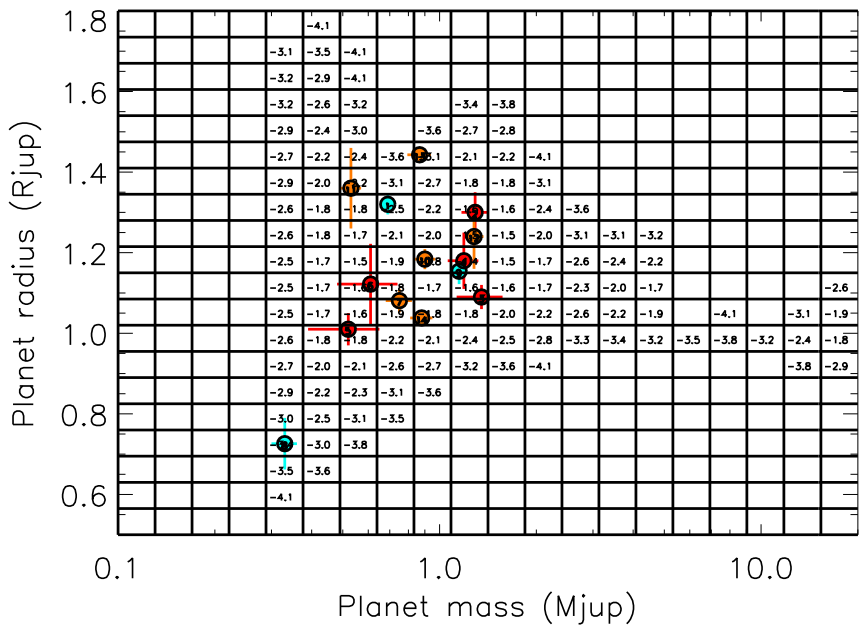}}}
\caption{ {\modif Logarithm of the probability that a simulated
detection event occurs in each one of the 20x20 bins of the mass/radius diagram. 
The likelihood of a multiple-events draw is the sum of the logarithms of the probabilities 
of the events of this draw. Bins without any occuring event in the large Monte-Carlo draw do 
not have any probability stated.
The likelihood of a $n$-events draw is the sum of the probabilities of its $n$ events.
In this mass-radius diagram, OGLE planets are shown as red circles, 
planets from other surveys are in orange, and planets from radial velocity surveys are in blue.
The likelihood of the 5 OGLE discoveries as
a result of a Monte-Carlo draw is $-8.7$, the maximum likelihood is $-7$ and the 
standard deviation to maximum likelihood is $2.54$. Hence, the result of the OGLE planets mass-radius
distribution is at $0.67 \sigma$ of the maximum likelihood of the model.
}}
\label{fig:m_r_probabilities}
\end{figure*}

Figure~\ref{fig:m_r_probabilities} shows the logarithm of the probability that an event occurs in each of the 20x20 bins of the mass-radius diagram. 
The likelihood of a draw of several independant events is defined as the sum of
the logarithms of the probabilities of these events. In order to compare our results to any $n$ real discoveries, we first estimate the standard deviation of any $n$-planets-random-draw 
compared to the maximum likelihood of the model. We randomly select $n$ planets among the simulated detections and calculate the likelihood of this draw. We repeat this selection $1000$ times in order to have the maximum likelihood and its standard deviation $\sigma$, then we compare the deviation of the likelihood of the $n$ real detecions calculated the same way in terms of $\sigma$. Henceforth, quantitative comparisons between the simulation results and the known planets are systematically given in the figure captions, whilst the text discusses qualitative comparisons and their implications.
For the different figures showing the results of our simulation, we compare the distribution of planets over the detection threshold to the 5 OGLE planets. We also compare our results to the 11 planets discovered by all transit surveys, as their detection biases are similar to OGLE, and to the 14 planets which radius is known (11 from transits and 3 from radial velocity surveys) to show how our model can reproduce the whole known population.}

\subsection{Depth of the transit events and magnitude of the targets stars}

\begin{figure*}
\vspace{0.52cm}
\centerline{\resizebox{13cm}{!}{\includegraphics{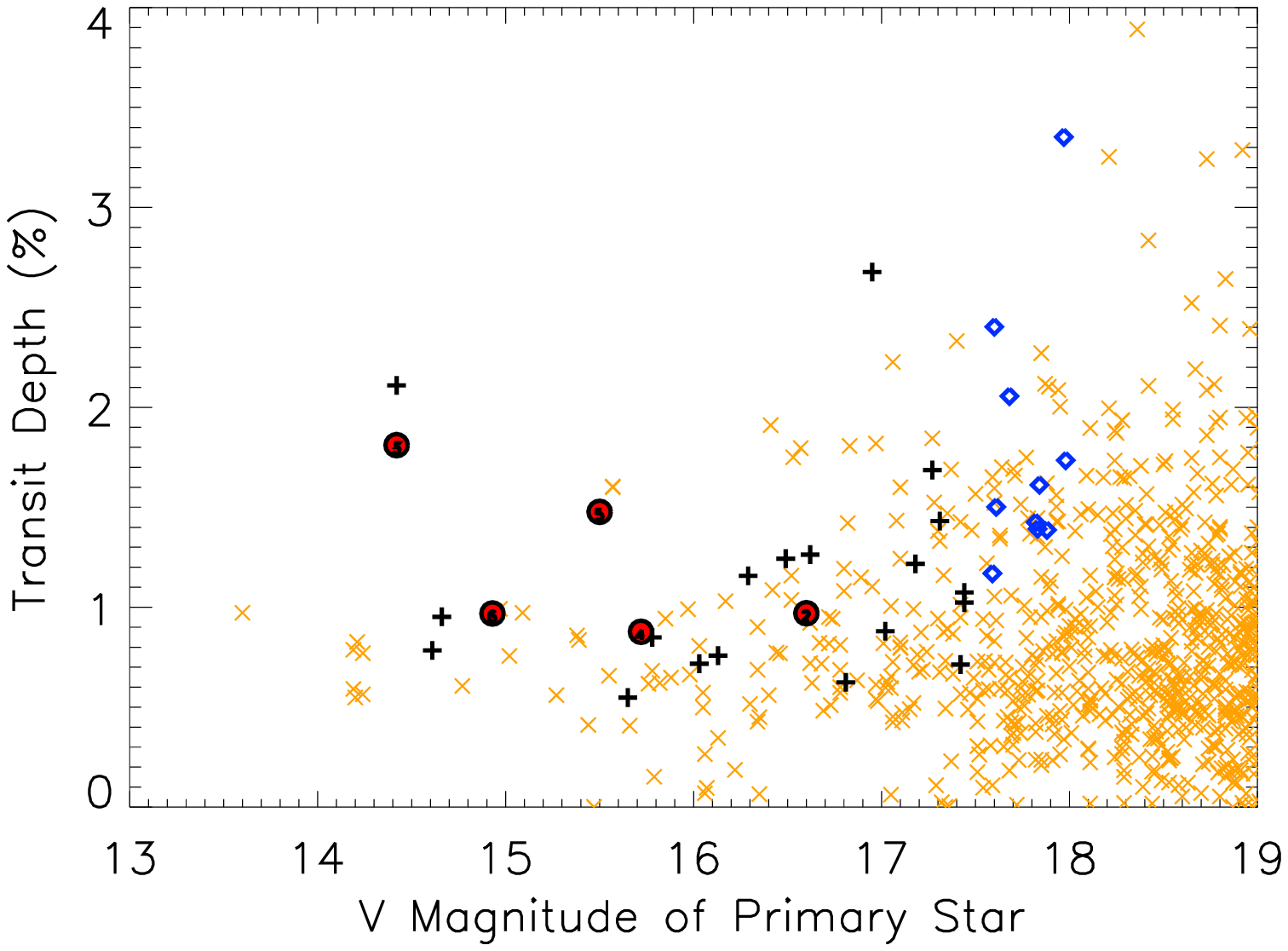}}}
\caption{Depth of the planetary transit events versus magnitude of the
  parent star in the V band. The five confirmed OGLE detections are shown
  as circles. {\modif Model results are shown as black plusses for detectable
  events and orange crosses for events that are considered
  undetectable based on the photometric signal (see text). Blue diamonds 
  correspond to events that would be detectable by photometry
  alone but that cannot be confirmed by radial velocimetry}. Note that
  the model results correspond to $3$ times the full OGLE campaign
  for more statistical significance. The OGLE planets depth-magnitude 
  distribution is at $0.69 \sigma$ from the maximum likelihood of the model.}
\label{fig:depth_magn_v19}
\end{figure*}

We first attempt to confirm whether the events detected by the model
are consistent with those found in the OGLE
fields. Figure~\ref{fig:depth_magn_v19} is a plot showing
transit depth as a function of the magnitude of the primary
star. {\modif Model results are considered detected when the signal-to-noise ratio
is sufficient for a detection (see \S~\ref{sec:noise levels}). 
We also show events that are considered photometrically
detectable but very hard or impossible to confirm by radial
velocimetry.}

The figure evidently shows a good correlation between the black
crosses and the red circles that indicate real detections by OGLE,
with a range of transit depths and V magnitudes that is very similar
between the models and the observations. Our models overpredict
slightly the number of transit events around faint stars ($V\ge 17$),
but this may be due to the difficulty of the follow-up work for these
targets. Overall, the agreement between models and observations is
good.

\subsection{Compatibility of transit surveys with radial-velocimetry
  observations} 
\label{sec:compatibility}

\subsubsection{Compatibility in the mass-period diagram}

\begin{figure*}
\vspace{0.52cm}
\centerline{\resizebox{13cm}{!}{\includegraphics{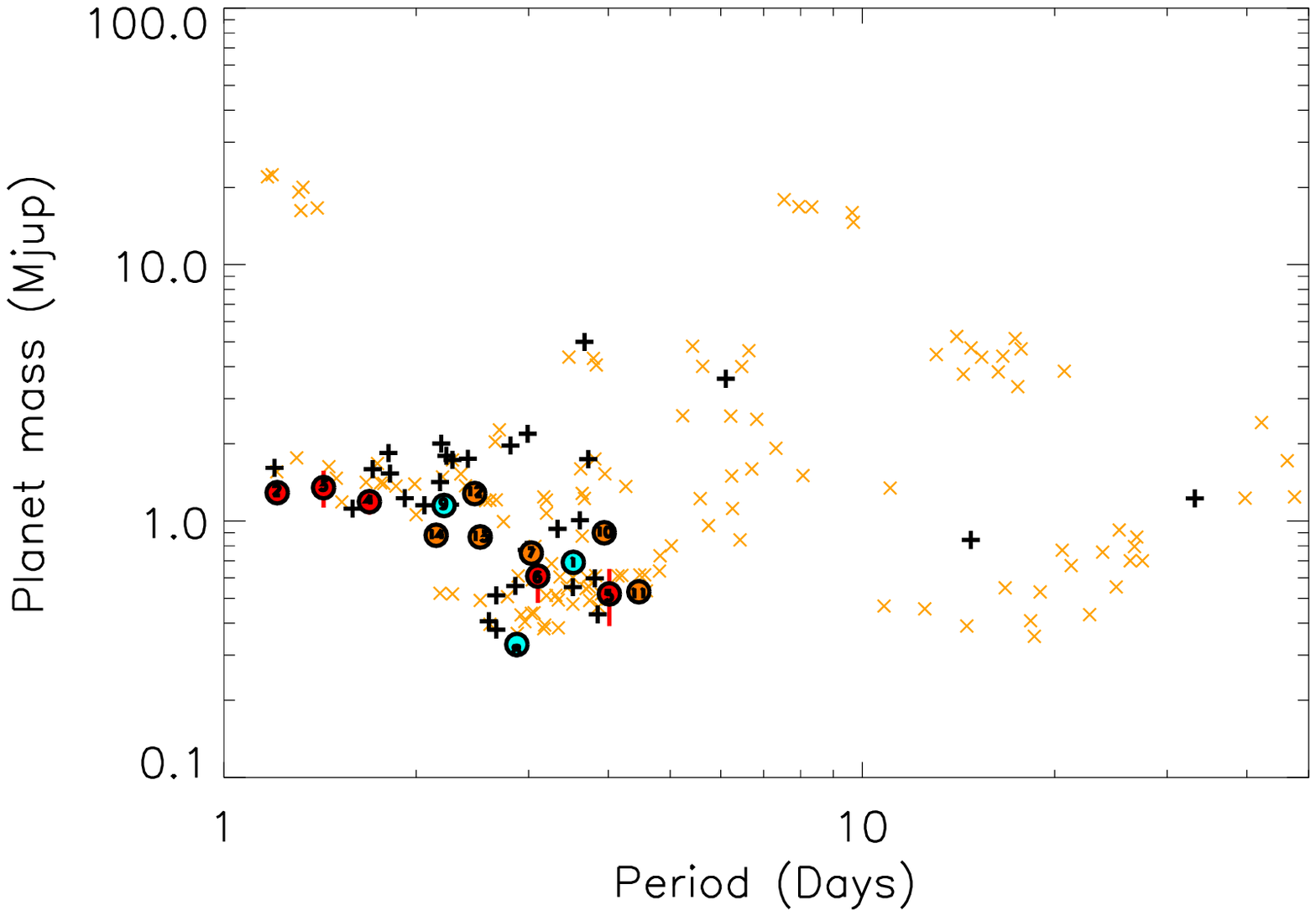}}}
\caption{Mass versus period of transiting giant planets.
{\modif (OGLE planets are red circles, other transit surveys in orange, planets from
radial velocity surveys in blue. Simulated planets detected: black plusses, under threshold: orange crosses).
The OGLE planets mass-period distribution is at $0.62 \sigma$ 
  from the maximum likelihood of the model ($0.72 \sigma$ considering 
  the 11 planets discovered by transit surveys and $0.66 \sigma$ considering the 14 known planets).
  }}
\label{fig:mass_period}
\end{figure*}

Figure~\ref{fig:mass_period} compares the model and observated
mass-period relation. As it is independant of the planetary evolution
model, it is a direct test of the compatibility between the results of
transit surveys and those of radial-velocimetry observations that
drive our model results. Again, the comparison is very good, assuming
a high-enough frequency of very-close in planets (see discussion
in \S~\ref{sec:statistics}). One can note especially the absence of
planets of relatively large mass (several times that of Jupiter) at
short orbital distances ($P<5\,$days), and of detectable transit
events for periods longer than $\sim 5$\,days. This is due especially to 
the fact that only events with a relatively large number of observed transits are detectable, as  
required by the $S_r$ threshold, which, given the day/night  
interruptions, imposes a constraint of a short orbital period. Note  
that this feature is not well reproduced by models in which the threshold
is computed from white-noise only \citep{Gould_2006, Gillon_2005}.

\subsubsection{The OGLE yields with a fixed red noise level}

\begin{table*}
\caption{OGLE yields with fixed red noise level}
\label{table:ogle_results}
\centering
\begin{tabular}{ccccccccr}
\hline\hline
\multicolumn2{c}{Field}  & Mean red & RV follow-up & \multicolumn4{c}{Number of planets}\\
\multicolumn2{c}{of view} & noise level & to Vmag &detected& \multicolumn3{c}{simulated with}\\
& & & & & 0 & 1.5 & 3\\
& & & & & \multicolumn3{c}{VHJ added ($P < 2$\,days)}\\\hline
\multicolumn2{c}{Bulge} & 3.6 & 17.5 & 2 & 0.4 & 0.6 & 0.9\\
\hline
Carina & original & 3.1 & 17.5 & 3 & 3.4 & 4.1 & 4.8\\
& updated & 2.1 & 17.5 & $+(0-1)$ & $+1.1$ & $+1.1$ & $+1.1$\\
\hline
\multicolumn2{c}{Centaurus}  & 3.1 & {\modif 17.0} & 0 & 1.4 & 1.8 & 2.2\\
\hline
\multicolumn2{c}{Total}  & & & 6 & 6.3 & 7.6 & 9.0\\
\hline\hline
\end{tabular}
\end{table*}

We have tested the efficiency of the fiducial model at estimating quantitatively the yield of transit surveys. \citet{Gillon_2005} have also simulated OGLE yield in their generic study of multiple transit surveys, but with restrictive assumptions on transit detectability (only complete events matter for detection purposes) and without considering background stars and red noise, also not using OGLE-fields specific stellar population. We also included in our simulations the recent {\modif RV follow-up} that has been done on Centaurus and Carina. 
We use unpublished information from  
the OGLE/ESO follow-up team, who found one promising planetary  
candidate among the Carina fields reprocessed with the systematics- 
removal algorithm from \citet{Tamuz_2005} and none in the Centarus  
fields, with a magnitude limit near V=17 for  the radial velocity  
follow-up. Table~\ref{table:ogle_results} compares the average number of planets 
detected for 1000 Monte-Carlo draws to real detections from the OGLE survey. 

The total number of simulated discoveries obtained from this quantitative analysis is in good agreement with the real detections. The differences in the number of detections between the Carina and Centaurus surveys are mainly due to the lower duty cycle of the observations towards Centaurus. A red noise level fixed at $3.6$ mmag in the direction of the galactic bulge bans most hot Jupiter detections.
The agreement between our quantitative result and the number of real detections is an indicator of the global efficiency of our approach (stellar and planetary distributions, evolution model and noise budget) for estimating transit survey yield.

\subsubsection{The OGLE yields with a variable red noise level}

So far, we have considered the level of red noise to depend only on the field considered. We attempt now to refine this by considering how the stellar density may affect it. 
Whereas most ground-based transit surveys have a global red noise level from $2$ to $3.5$ mmag (Superwasp: \citet{Smith_2006}, Monitor: \citet{Irwin_2007}, Hatnet: Pont $\&$ ISSI team (2007) and OGLE), the causes of these noise levels seem different, with instruments ranging from 10-cm wide field reflectors to deep-sky several-meter telescopes. As seen from table~\ref{table:ogle_results}, the OGLE fields have different mean red noise levels ($\sigma_r=3.6$\,mmag for the bulge and $\sigma_r=3.1$\,mmag for Centaurus and Carina before SYS-REM), although the instrument and observational strategy were unchanged. 
Looking at what distinguishes these fields, it appears that the most significant difference is the stellar density and therefore the amount of crowding: The bulge field is about twice as dense as the Carina and Centaurus fields. Pont $\&$ ISSI team (2007) {\modif raise the suspicion that the level of red noise depends strongly on} the presence and characteristics of contaminating stars, because e.g. of their different colors and differential refraction in the atmosphere. It is hence natural to consider a red noise that depends on a crowding index. 

We define this crowding index as the fraction of the flux coming from background stars versus that from the target {\modif in the photometric aperture}. Importantly, we do not consider stellar companions as contributing to the red noise because they are generally on the same CCD pixel as the target star and should affect {\modif the noise budget much less}. 

\begin{figure}
\vspace{0.52cm}
\centerline{\resizebox{1\linewidth}{!}{\includegraphics{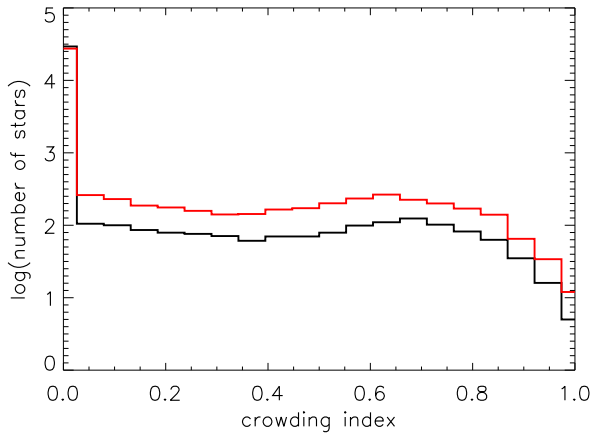}}}
\caption{Distribution of the crowding index (see text) of \modif{target stars} in Carina (black) and in the bulge (red).}
\label{fig:crowding_bulge_carina}
\end{figure}

Figure~\ref{fig:crowding_bulge_carina} shows the differences of crowding index for the target 
stars with planetary transits (detectable or not) in simulations of the Carina and Bulge fields of view. 
The mean crowding index for target stars of {\modif $I<17$} is $0.11$ in the Carina field and $0.233$ in the Bulge field.

We can exclude the fact that all red noise is linked
with contamination as many stars {\modif in the Carina} fields are unblended by background stars but
still show a high noise level. 

In order to estimate of the influence of the crowding on the red noise level, we use the following simple relation between red noise level and crowding index: 
\begin{equation}
\sigma_{r}=\alpha \times F_{b} + \beta,
\end{equation}
where $F_{b}$ is the fraction of total flux from background stars, determined on a star-by-star basis in our simulations, and $\alpha$ and $\beta$ are parameters to be determined. This is  
justified by the behaviour of the red noise seen for instance in 
SuperWASP, {\modif showing a linear increase as a function of background flux} \citep{Smith_2006}.
In order to get the same mean red noise values as \citet{Pont_2006}, we 
obtain $\alpha = 0.4$\,mmag and $\beta=2.65$\,mmag. This value of $\beta$ corresponds to the minimum red noise level obtained for non-contaminated stars in the OGLE fields. 

Table~\ref{table:ogle_crowding_results} shows the new number of detections when
considering this crowding-dependant red noise level. Compared to 
table~\ref{table:ogle_results}, the number of detections is found to be 
essentially unchanged for the Carina and Centaurus fields, 
but it increases by a factor $\sim 3$ for the bulge field. 
This result is more satisfactory because in the previous case, 
only $\sim 5$\% of the simulations would yield the detection of 
2 planets in the bulge, as observed.

\begin{table*}
\caption{OGLE yields with variable red noise level}
\label{table:ogle_crowding_results}
\centering
\begin{tabular}{cccccccr}
\hline\hline
\multicolumn2{c}{Field}  & RV follow-up & \multicolumn4{c}{Number of planets}\\
\multicolumn2{c}{of view} & to Vmag &detected& \multicolumn3{c}{simulated with}\\
& & & & 0 & 1.5 & 3\\
& & & & \multicolumn3{c}{VHJ added $(P < 2 days)$}\\
\hline
\multicolumn2{c}{Bulge}     & 17.5 & 2 & 1.2 & 1.6 & 2&\\
\hline
Carina & original    & 17.5 & 3 & 3.6 & 4.3 & 4.9 & \\
& updated & 17.5 & $+(0-1)$ & $+1.1$ & $+1.1$ & $+1.1$ &\\
\hline
\multicolumn2{c}{Centaurus} & 17 & 0 & 1.3 & 1.9 & 2.3&\\
\hline
\multicolumn2{c}{Total} & & 5-6 & 7.2&8.9&10.3&\\
\hline\hline
\end{tabular}
\end{table*}

\subsubsection{Models, observations and the frequency of very close-in planets}
\label{sec:statistics}

As discussed in \S~\ref{sec:probability}, three OGLE planets have
orbital periods shorter than 2 days and thus belong to a class of
objects yet to be detected by radial velocimetry. So far, we have
added one such planet (on average) to our carbon copy list of nearly
200 radial velocimetry planets. In Section~\ref{sec:compatibility}, we
have shown that with this assumption, radial-velocity and photometric
transit surveys are compatible. We now test the range of
frequencies of very close-in planets for which this remains true. 

\begin{figure}
\vspace{0.52cm}
\centerline{\resizebox{1\linewidth}{!}{\includegraphics{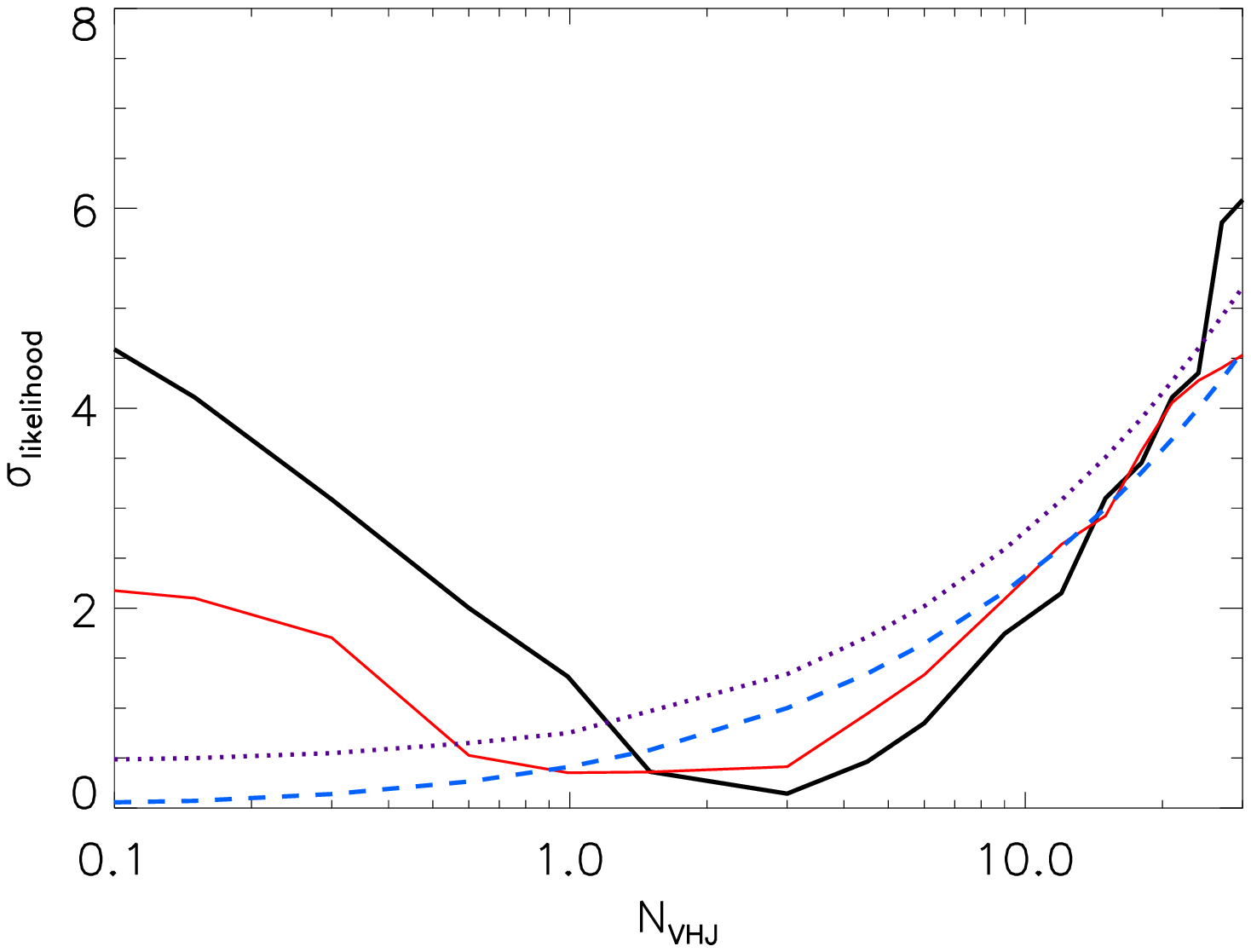}}}
\caption{Deviations from a maximum likelihood obtained as a function
  of $N_{VHJ}$, the number of very hot jupiter of orbital periods
  shorter than 2 days added to the radial velocities carbon-copy list.
  {\it Thick line}: Deviation from the maximum likelihood obtained in
  the mass-radius diagram for the OGLE planets. {\it Thin line}:
  Same deviation but when compared to the ensemble of planets. {\it
  Dashed line}: Standard deviation obtained from a comparison between
  the number of simulated planets and the number of detected ones for
  the OGLE survey (see table~\ref{table:ogle_crowding_results}). {\it
  Dotted line}: Standard deviation obtained from the non-detection of
  these very close-in planets by radial-velocimetry.}
\label{fig:VHJ}
\end{figure}

In order to do so, we compute the deviation from maximum likelihood in
the mass-radius diagram like in Section~\ref{sec:compatibility}, as a
function of the number of planets {\modif which period is less than $2$ days} added to the RV list. The
result is presented in Fig.~\ref{fig:VHJ} and shows that a good match
is obtained by adding 1 to 3 short-period planets. Larger numbers are
also possible from the point of view of the transit surveys, but would
conflict with their non-detection by radial-velocimetry. Adding the
other transiting planets discovered thus far yields smaller
probabilities of occurence of these short-period planets, but not
by significant amounts.

All in all, and assuming that the radial velocity planets sample is
unbiased, we constrain the fraction of main-sequence late stars
orbited by very hot giant planets with orbital periods less than 2
days to be 
$(1/1265)(1_{-0.33}^{+0.33})$ at a 60 \% confidence level or
$(1/1265)(1_{-0.5}^{+0.83})$ at a 90 \% confidence level.

The distribution of planets in period between 2 and 5 days is directly
obtained from the metallicity-linked distribution \citep{Santos_2004}
and the RV planets sample.  Adding the distribution we found for
planets between 1 and 2 days, we obtain a fraction of $(1/215)$ late
main-sequence stars orbited by planets in the 1 to 5 days period
range, in good agreement with the results obtained in
\citet{Gould_2006}, who obtained $(1/220)(1^{+1.10}_{-0.45})$.
Similarly, the distribution we obtain by cutting this sample into two
parts with the cut-off at 3 days is compatible, showing:

\begin{itemize}
\item a slightly higher fraction of really short-period planets (1-3
days) of $(1/560)$ 
{\modif instead of} $(1/710)(1^{+1.10}_{-0.54})$ at a 90 \% confidence level in
\citet{Gould_2006}.
\item a similar fraction of short-period planets (3-5 days) of
$(1/350)$ 
{\modif instead of} $(1/320)(1^{+1.39}_{-0.59})$ at a 90 \% confidence level in
\citet{Gould_2006}.
\end{itemize}



The results presented hereafter use the variable red noise level
approach, and an RV planet list that is complemented with, on average,
1.5 very-close in planets with periods $P<2$\,days taken from the OGLE
detections. 

\subsection{The metallicity of the stars harboring transiting planets}
\label{sec:metallicity}

\begin{figure*}
\vspace{0.52cm}
\centerline{\resizebox{13cm}{!}{\includegraphics{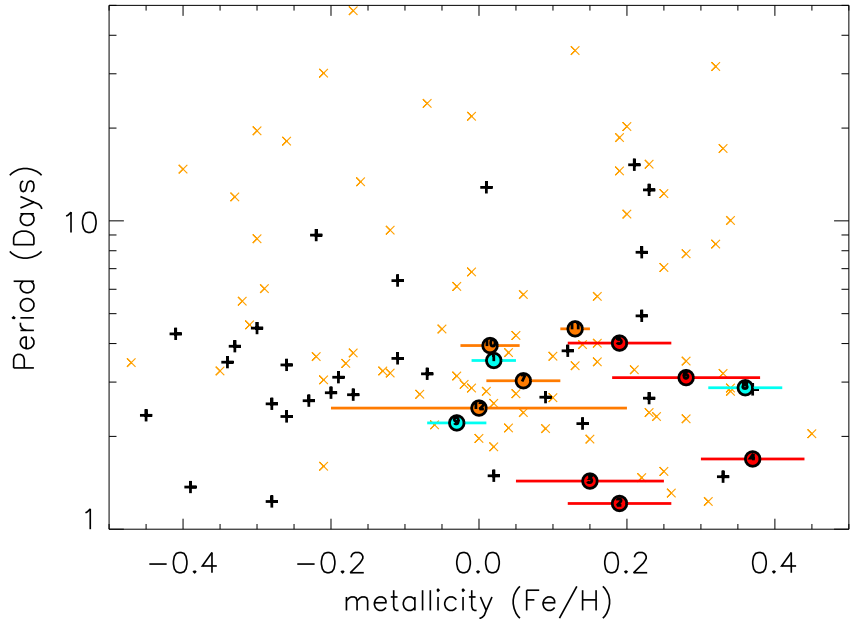}}}
\caption{Period of transiting exoplanets versus metallicity of their
  parent star. The model is based on analytic relations for the mass
  and period distributions of planetary companions (see
  \S~\ref{sec:masses and orbits}). {\modif (OGLE planets are red circles, other transit surveys in orange, planets from radial velocity surveys in blue. Simulated planets detected: black plusses, under threshold: orange crosses).
  The OGLE planets period-metallicity
  distribution is at $2.94 \sigma$ from the maximum likelihood of the model
($2.51 \sigma$ considering the 11 planets discovered by transit surveys and $2.63 \sigma$ considering the 14 known planets).
  }}
\label{fig:period_metallicity_no_cut}
\end{figure*}

We now compare the metallicity of the parent stars for our observed
and modelled populations. A first test using the analytical scenario
for the radial-velocity population
(Fig.~\ref{fig:period_metallicity_no_cut}) yields a clearly different
metallicity distribution, with most of the transiting planets observed
around low-metallicity stars. We verified that this problem occurs
independantly of the assumed stellar metallicity distribution, for any
realistic stellar population. It arises fundamentally because the
global metallicity bias as obtained by \citet{Santos_2004} or \citet{Fischer_valenti_2005} 
is not strong enough to compensate for the rarity of
very metal-rich stars in the Galaxy.

\begin{figure*}
\vspace{0.52cm}
\centerline{\resizebox{13cm}{!}{\includegraphics{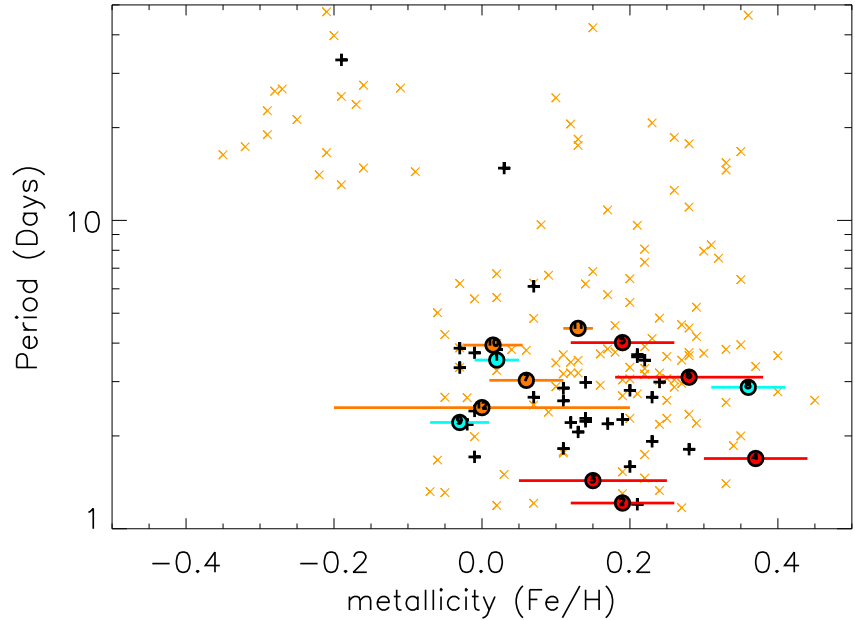}}}
\caption{Period of transiting exoplanets versus metallicity of their
  parent star. The figure differs from
  Fig.~\ref{fig:period_metallicity_no_cut} in that our fiducial model,
  i.e. the mass-period-metallicity ``carbon-copy'' model is used (see
  \S~\ref{sec:masses and orbits}). 
  {\modif (OGLE planets are red circles, other transit surveys in orange, planets from
  radial velocitiy surveys in
  blue. Simulated planets detected: black plusses, under threshold: orange crosses).
  The OGLE planets period-metallicity
  distribution is at $0.76 \sigma$ from the maximum likelihood of the model
 ($0.36 \sigma$ considering 
  the 11 planets discovered by transit surveys and $0.39 \sigma$ considering the 14 known planets)
  .}}
\label{fig:period_metallicity}
\end{figure*}

As seen in Fig.~\ref{fig:period_metallicity}, the problem disappears
when one considers the carbon-copy model. Thus, we are led to an
important conclusion, that the metallicity distribution of pegasids
(periods shorter than 10 days) is fundamentally different from the
global exoplanet population. More specifically, there are no {\modif Pegasids} 
orbiting F, G, K stars with metallicities smaller than
[Fe/H]$=-0.07$. This has
strong consequences for planet formation models \citep[see also][]{Guillot_2006}. 
This work shows that this conclusion is robust, and is
needed to explain the results of the photometric surveys.

A finer examination of Fig.~\ref{fig:period_metallicity} shows that
while our model planets reproduce globally the metallicity of the
ensemble of transiting planets, OGLE stars with planets are on average
$\sim 0.1$\,dex more metal-rich. 

This can tentatively be explained with a metallicity gradient in the
galaxy for OGLE TR-10 ($\rm [Fe/H]=0.28\pm{0.10}$) and OGLE TR-56
($\rm[Fe/H]=0.19\pm{0.07}$), the two planets discovered in the direction
of the galactic bulge.  The study of galactic cepheids by
\citet{Andrievsky_2004} shows a metallicity gradient as a function of
distance to the galactic center.  In the $[6.6,10.6]$ kpc-range
distance from galactic center, this study finds a linear relation
between $[Fe/H]$ and galactocentric distance $R_G$:
\begin {equation}
\rm[Fe/H]=-0.044(\pm {0.004}) R_G + 0.363(\pm {0.032}) 
\label{eq:[Fe/H]_R_G}
\end {equation}
Following that relation, the two stars with planets discovered in the
direction of the galactic bulge both at a distance around 1500 pc
would thus be in a $0.04$ dex more metal rich region than the solar
neighborhood.

Concerning the high metallicity of stars with transiting planets
discovered by OGLE in the Carina region, we do not have any reason to
think that the metallicity distribution would be different from the
solar neighborhood. Our only hypothesis is a low-probability draw for
metallicity for the 3 OGLE-Carina planets.

\subsection{Atmospheric potential energy and orbital distances}

\begin{figure*}
\vspace{0.52cm}
\centerline{\resizebox{13cm}{!}{\includegraphics{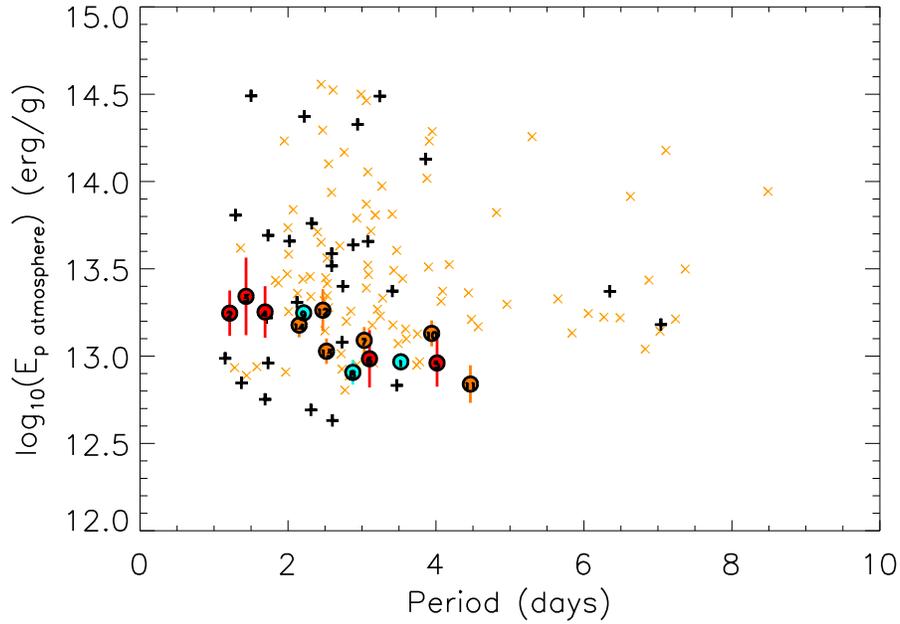}}}
\caption{Potential energy per unit mass ($E_p=GM/R$) versus orbital period of
  transiting planets. {\modif (OGLE planets are red circles, other transit surveys in orange, planets from radial velocity surveys in 
  blue. Simulated planets detected: black plusses, under threshold: orange crosses).
  Observations are compared to models based on the
  analytical relations for the mass and period distribution of
  planetary companions (see \S~\ref{sec:masses and orbits}).
  The OGLE planets energy-period distribution is at
  $2.18 \sigma$ from the maximum likelihood of the model ($1.86 \sigma$ considering 
  the 11 planets discovered by transit surveys and $2.47 \sigma$ considering the 14 known planets).
}}
\label{fig:nrj_period_analytical}
\end{figure*}

Because evaporation may affect the planet population, it is
instructive to check whether the potential energy of the atmosphere
and the orbital period, two crucial quantities for this process
\citep[e.g.][]{LdEVMH04}, also possess a relatively consistent
distribution. We first test the behavior of the analytical model for
the distribution of planets
(Fig.~\ref{fig:nrj_period_analytical}). This results in a prediction
of many planets with large radii (small values of the potential energy
for atmospheric escape) at small orbital distances, in patent
contradiction with the observations. 

\begin{figure*}
\vspace{0.52cm}
\centerline{\resizebox{13cm}{!}{\includegraphics{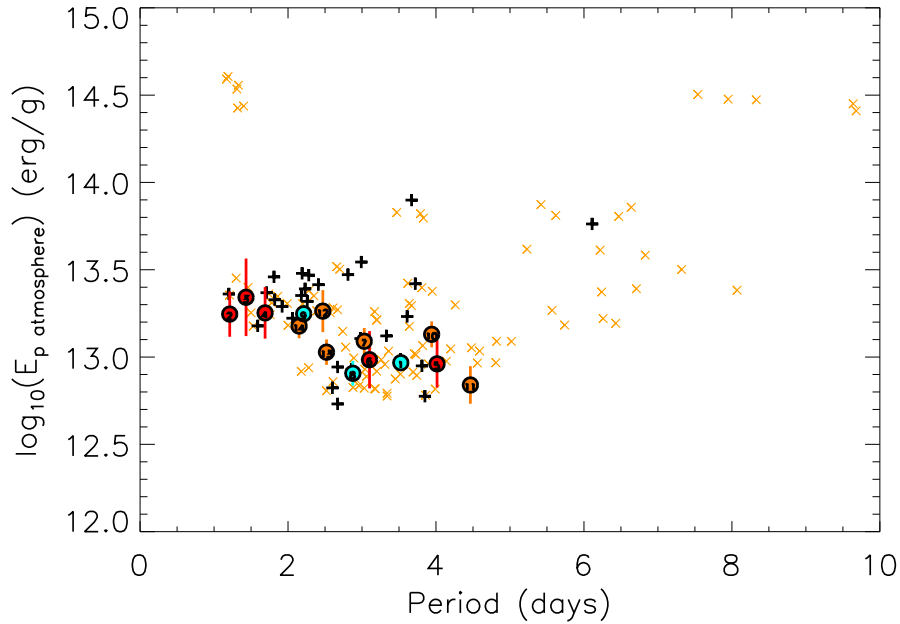}}}
\caption{Potential energy per unit mass versus orbital period of
  transiting planets. The figure is similar to
  Fig.~\ref{fig:nrj_period_analytical}, except for the fact that our
  fiducial model is used (see \S~\ref{sec:masses and orbits}). 
   {\modif (OGLE planets are red circles, other transit surveys in orange, planets from radial velocity surveys in
  blue. Simulated planets detected: black plusses, under threshold: orange crosses).
  The OGLE planets energy-period distribution is at
  $0.55 \sigma$ from the maximum likelihood of the model. ($0.84 \sigma$ considering 
  the 11 planets discovered by transit surveys and $0.66 \sigma$ considering the 14 known planets)
  }}
\label{fig:nrj_period}
\end{figure*}

The problem mostly disappears with the carbon-copy model:
Fig.~\ref{fig:nrj_period}) shows that in this case, although we do not
obtain a linear correlation between the two variables, we get
detections in the right area of the diagram. This is explained as
stemming from:
\begin{itemize}
\item The absence of low-mass planets at small orbital distances, with
  a possible limiting relation between these two quantities \citep{Mazeh_2005};
\item The difficulty in detecting planets with larger values of \modif{potential energy per unit mass}
  (smaller radii) at large orbital distances --although we predict
  that some of these should be detected by future transit surveys.
\end{itemize}

Our results strengthen the case for the existence of a relation between
mass and orbital distance for short-period planets, as advocated by
\citet{Mazeh_2005}: Indeed, the analytic model which is characterized
by the presence of small mass planets at small distances yields a
distribution of detectable planets that is significantly different
from the observations (Fig.~\ref{fig:nrj_period_analytical}). 
Our carbon-copy model that includes implicitely this correlation does
not (Fig.~\ref{fig:nrj_period}).

\subsection{Planetary radii and stellar irradiation}

\begin{figure*}
\vspace{0.52cm}
\centerline{\resizebox{13cm}{!}{\includegraphics{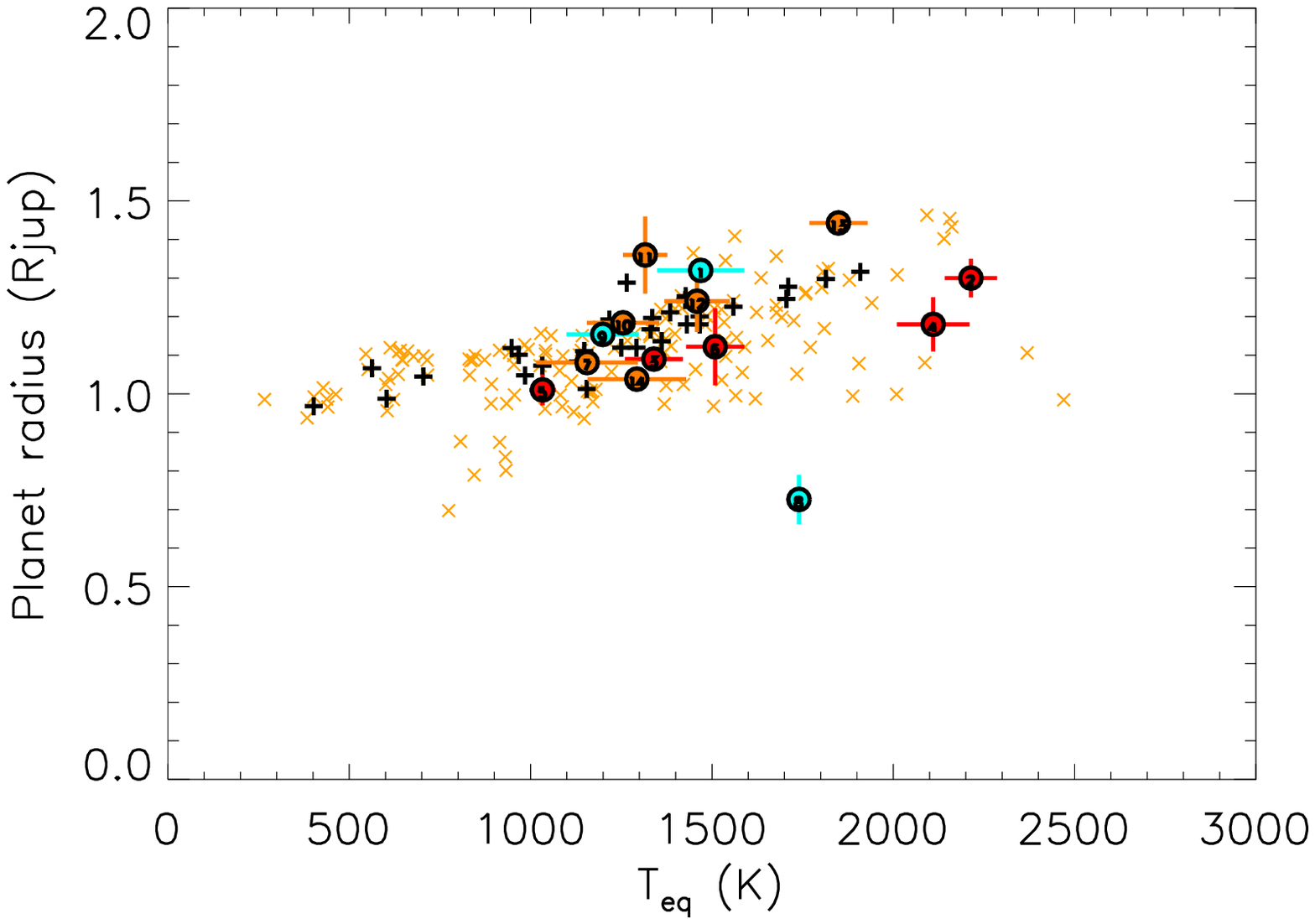}}}
\caption{Radius as a function of equilibrium temperature of transiting
  exoplanets. 
   {\modif (OGLE planets are red circles, other transit surveys in orange, planets from radial velocity surveys in 
  blue. Simulated planets detected: black plusses, under threshold: orange crosses).
  The OGLE planets 
  equilibrium temperature-radius distribution is at $1.22 \sigma$ from the maximum 
  likelihood of the model ($1.05 \sigma$ considering 
  the 11 planets discovered by transit surveys and $2.25 \sigma$ considering the 14 known planets).}}
\label{fig:teq_radius}
\end{figure*}

Radius and stellar irradiation (or equivalently equilibrium
temperature) should be positively correlated, as a planet with a
higher irradiation dose will tend to cool and contract more slowly
than one that endures less stellar insolation. As
Fig.~\ref{fig:teq_radius} shows, the correlation exist, but is weak,
and with a signficant scatter. This is well reproduced by the
model. 

However, it can be noted that HD~149026~b lies away from the cloud of
points. In general, we find that our fiducial model generates few
points in this region. This can be easily accounted for by slightly
modifying the metallicity-core mass relation to allow for larger
masses. As planets of small masses and large core masses are more
difficult to model anyway, we chose not to attempt fine-tuning the
model to this level of detail. This should be postponed for further
studies, especially with the discovery of more Saturn-mass transiting
planets.

\subsection{The mass-radius relation}

We have checked that our fiducial model predicts the detection of
transiting planets with properties that are globally consistent with
the observations. We can now examine in more detail the mass-radius
relation thus obtained, as it is directly tied to assumptions on the
compositions and evolutionary models of exoplanets. The predictions
also have implications for transit surveys as it is not clear
whether they have detected only the ``tip of the iceberg'', ie the few
largest giant planets while many smaller ones would lie undetected or
not. 

\begin{figure*}
\vspace{0.52cm}
\centerline{\resizebox{13cm}{!}{\includegraphics{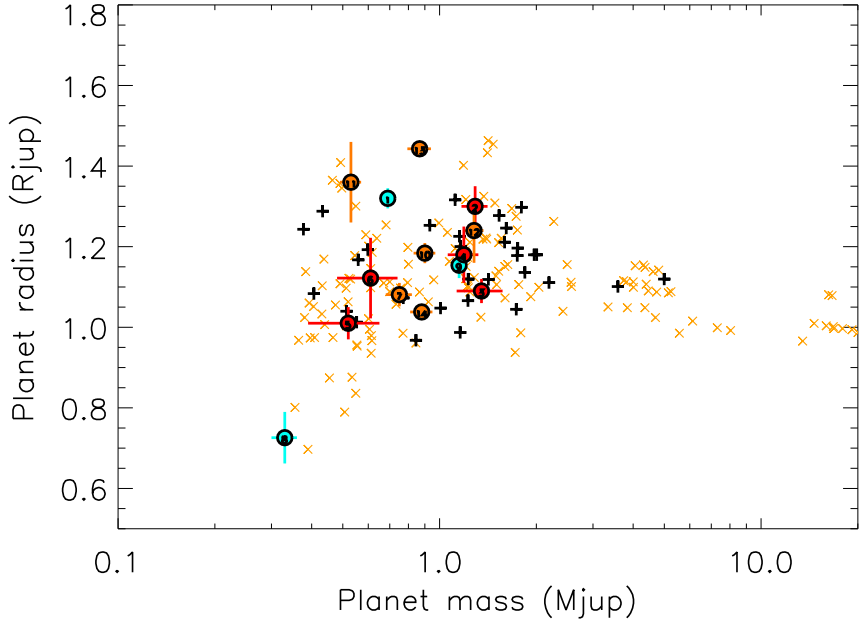}}}
\caption{Mass-radius relation for transiting extrasolar giant
  planets. 
{\modif (OGLE planets are red circles, other transit surveys in orange, planets from radial velocity surveys
  in blue. Simulated planets detected: black plusses, under threshold: orange crosses).
  The OGLE planets mass-radius distribution is at 0.67
  $\sigma$ from the maximum likelihood of the model ($0.72 \sigma$ considering 
  the 11 planets discovered by transit surveys and $0.97 \sigma$ considering the 14 known planets)
  .}}
\label{fig:radius_mass}
\end{figure*}

Results with our fiducial model are presented in
Fig.~\ref{fig:radius_mass}. We find that planets with low masses (say,
less than Jupiter's mass) can both have very large or very small
radii, depending on whether they contain a significant mass in heavy
elements or not. On the other hand, {\modif massive} planets have radii which are
comparatively better defined. This is mostly due to the fact that we
assume a maximum mass of heavy elements of 100\,M$_\oplus$, a
hypothesis that will be tested directly by the discovery of a few
massive transiting planets.

Our model results once again agree well with the detections made by
photometry. Importantly, the yellow crosses in
Fig.~\ref{fig:radius_mass} do not lie significantly below the black
ones: we predict that future surveys will {\it not} discover a population of
small-sized giant planets, at least for masses above that of
Saturn. 

The presence of planets with larger masses of heavy
elements should remain marginal because otherwise they would have
been detected by present-day surveys, Fig.~\ref{fig:radius_mass}
showing that planets below 1\,R$_{\rm Jup}$ are already
detectable, although in favorable cases (small radius of the primary
and bright targets). Quantitatively, simulations in the OGLE fields
indicate that if planets had radii uniformly distributed between 0.5
and 1.5 \,R$_{\rm Jup}$, 18.5\% of the planets discovered by the survey
would have radii below 1\,R$_{\rm Jup}$. This fraction is not
negligible and is (although marginally) inconsistent with the sample
of 0/11 planets with $R<R_{\rm Jup}$ discovered by transit surveys
thus far. 

Therefore, although we cannot statistically rule out the presence of a
population of small planets, these would require the formation of
extremely metal-rich planets. Our prediction is a consequence
of evolution models and of our assumption that planets with masses of
heavy elements beyond 100\,M$_\oplus$ should be rare.

\begin{figure*}
\vspace{0.52cm}
\centerline{\resizebox{13cm}{!}{\includegraphics{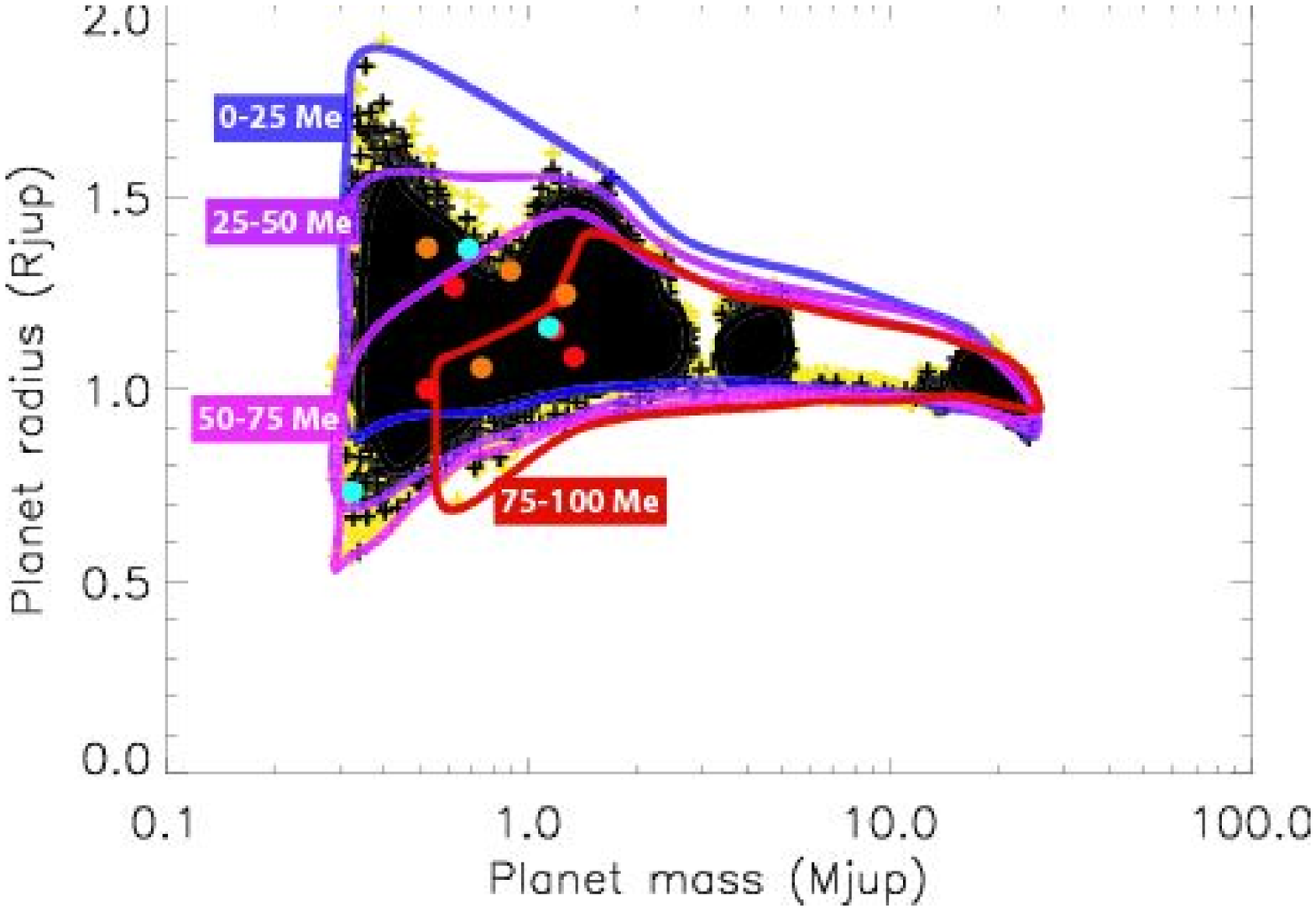}}}
\caption{Mass-radius relation for a very large number of Monte-Carlo
  trials using the fiducial model. The curves show the ensemble of
  planets with masses of heavy elements between 0 and 25, 25 and 50,
  50 and 75, 75 and 100\,M$_\oplus$, respectively. Symbols are as in
  Fig.~\ref{fig:mass_period}.}
\label{fig:gt43_mz}
\end{figure*}

Figure~\ref{fig:gt43_mz} shows the ensemble of planets obtained for an
extremely large number of draws, with our fiducial model. Voids in the
ensemble of crosses correspond to the absence of planets with these
masses in the radial-velocimetry list. They should not be considered
as significant. The contours in the figure indicate the ensemble of
masses and radii expected for planets with different masses of heavy
elements, from 0 to 100\,M$_\oplus$. Importantly, the location of
these contours is linked to our assumption of an energy source in the
planetary interior equal to 0.5\% times the irradiation received by
the planet. {\modif Independently of the details of this assumption, this shows that a statistically significant
ensemble of known transiting planets would allow a determination of
the presence or lack of heavy elements in these objects.}

\begin{figure*}
\vspace{0.52cm}
\centerline{\resizebox{13cm}{!}{\includegraphics{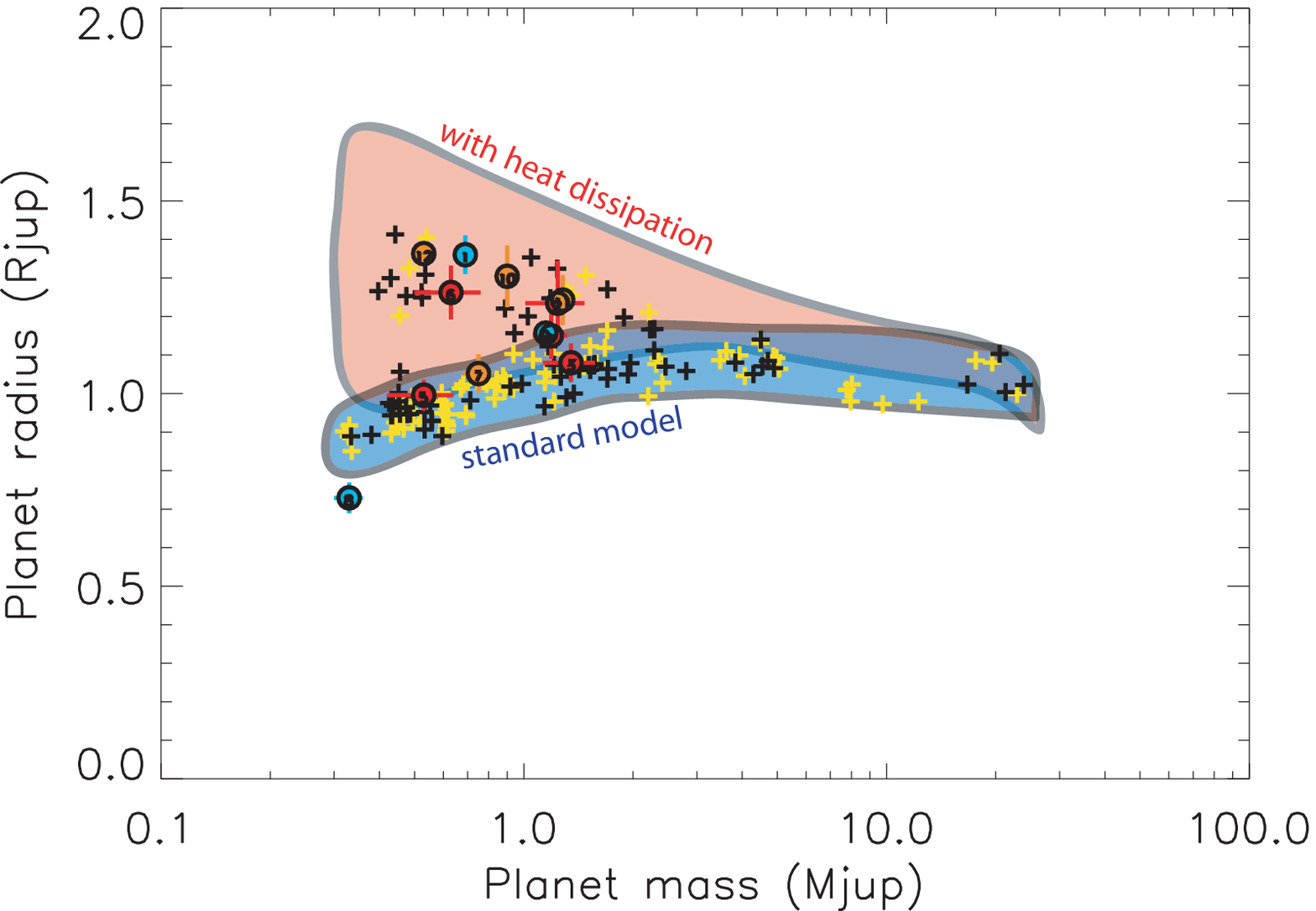}}}
\caption{Mass-radius relation obtained for an alternative model with
  70\% of ``standard'' planets with no extra-energy source, and 30\%
  planets receiving an additional $3\times 10^{26}\rm\,erg\,s^{-1}$
  luminosity dissipated at the center. All planets are assumed to
  possess $20\,\rm M_\oplus$ in heavy elements. Symbols are as in
  Fig.~\ref{fig:mass_period}. }
\label{fig:2bags_comp}
\end{figure*}

We have also tested another assumption regarding the planetary
evolution model: all planets possess 20\,M$_\oplus$ mass in heavy
elements, 70\% of them have no extra heat source, whereas 30\% have
$3\times 10^{26}\,\rm erg\,s^{-1}$ dissipated at the center. With this
assumption, one can qualitatively explain the observed transiting
planets (i.e. the ``normal'' planets and the ``anomalously large''
ones, respectively) with the exception of HD~149026~b, for which one
could argue that the planet comes from a different population. In this
case, Fig.~\ref{fig:2bags_comp} shows a distribution of radii that is
relatively similar to the previous one (Fig.~\ref{fig:radius_mass}),
with the exception that no planet has a radius smaller than $0.8\,\rm
R_{Jup}$. In this case, the 2 regions corresponding to the
``standard'' model, and to the ``heat dissipation'' case are clearly
different, especially at the low-mass range of the diagram.

Present observations cannot distinguish between the two
models, showing the need for additional detections of transiting giant
planets. Particularly important are planets between the mass of Saturn
and that of Jupiter, as this is a mass regime where expected
compositional differences have the largest impact.

\section{Conclusions}

We have presented a simulation of photometric transiting surveys based
on basic knowledge of the stellar and planetary populations in the
galactic neighborhood and on a planetary evolution model tuned to the
information obtained from transiting giant planets with masses above
that of Saturn. This simulation was applied to the OGLE survey, and shown
to yield a generally excellent agreement with the transiting planets
detected by the survey.

We have thus shown that radial velocimetry and photometric surveys are
compatible within statistical uncertainties, in agreement with
\citet{Gould_2006}. We have derived a frequency of very close-in
planets with orbital periods shorter than 2 days around solar-type
stars, of $(1/1265)(1_{-0.33}^{+0.33})$ at a 60 \% confidence level or
$(1/1265)(1_{-0.5}^{+0.83})$ at a 90 \% confidence level.

Using null results by photometric surveys for given ranges of
parameters, we are able to strengthen two results already present in
the radial velocimetry data:
\begin{itemize}
\item Stars with low metallicities ([Fe/H]$ < -0.07$) do not, or
  are very unlikely to harbour close-in giant planets with orbital periods
  $P < 10$ days. This is unlike stars above that metallicity
  threshold (see Fig.~\ref{fig:period_metallicity}). 
\item There is a lack of small-mass giant planets below the mass of
  Jupiter and above that of Saturn for orbital periods
  $P < 3$\,days (see Fig.~\ref{fig:mass_period}). 
\end{itemize}
Further data is required to {\modif precisely} quantify these empirical results that bear
important consequences for our understanding of planet formation and
migration. 

On the basis of our model, and assumptions concerning the composition
of giant planets (i.e. masses of heavy elements between 0 and
100\,M$_\oplus$), we find that the present detections of transiting
planets have sampled a population that is quite representative of the
main population of giant planets, at least for the ones that are above
about half the mass of Jupiter. We hence predict that future
transit surveys with higher sensitivities will {\it not}\/ discover a
significant population of yet undetected Jupiter-mass planets with small 
sizes, i.e. radii smaller than that of Jupiter (see
Fig.~\ref{fig:radius_mass}).

Many ground-based transit surveys are in progress, and with the space
missions CoRoT \citep{Baglin_2002} and Kepler \citep{Borucki_2003}, 
{\modif the number of known transiting planets is expected to rise rapidly over the next few years}. 
This will enable us to
better test the models and quantify some of the results presented in
this article. We also hope to be able to discriminate between various
models of the evolution and compositions of giant planets, a matter of
great importance for formation models. 

We wish to stress however that a continuation of ground-based transit
surveys is desirable even in the presence of similar programs from
space. CoRoT will survey 60,000 dwarf stars over five 150
days periods and Kepler about 100,000 over 4 years, implying a maximum
potential yield of 55 and 90 transiting giant planets, respectively,
plus many other smaller planets. For what concerns giant planets,
quantifying the fraction of very close-in planets with a 10\% accuracy
at the $3\,\sigma$ level would require the discovery of $\sim 200$
transiting planets. Understanding the evolution and compositions of
giant planets {\modif will require an even larger number of detections}. The radius of a giant
planet itself depends mainly on four parameters: the planetary mass,
equilibrium temperature, age, and its composition (note that the
composition can be considered as a simple parameter only in the case
of planets mostly made of hydrogen and helium: smaller planets will be
more difficult to model!). Additional energy
sources may occur (such as in the presence of tidal heat dissipation),
and the initial conditions and formation history may have their say in
the matter as well. Furthermore, the
observational uncertainties are generally large. For example, the
planetary radius is generally only known to $\sim 10\%$, for a global
variation that is relatively small (1 to 1.5\,R$_{\rm Jup}$). This
implies that to constrain a given correlation to, say 10\%, {\modif and 
with four independant variables, hundreds of data points are needed,
and thousands would be desirable.}

{\modif This motivates} us to
seek programs capable of detecting thousands of transiting planets in
the mid-term future, and ways to reduce the error bars on the
different parameters. One direction is to test the Dome C plateau in
Antarctica for such an ambitious program, which is the purpose of
A~STEP \citep{Fressin_2005}. Other directions exist, such as proposals
for similar surveys from space. In any case, it is most important that
a statistically significant population of exoplanets be characterized
for a better understanding of planet formation and our origins.

\section*{Acknowledgments}

The code used for this work, CoRoTlux, has been developped as part of
the CoRoT science program by the authors with major contributions by
Aur\'elien Garnier, Maxime Marmier, Martin Vannier, Suzanne Aigrain
and help from Claire Moutou, St\'ephane Lagarde, Antoine Llebaria,
Didier Queloz and Fran\c{c}ois Bouchy. We want to thank Andrzej
Udalski and Michael Gillon for their communications on OGLE data,
{\modif Fr\'ed\'eric Th\'evenin for his advices on stellar populations
simulation.}  F.F. has been funded by grants from the French {\it
Minist\`{e}re de la Recherche} and by the {\it Soci\'{e}t\'{e} des
Amis des Sciences}. V.M. was funded by a grant from the {\it
C.N.R.S.}. This work has used extensively Jean Schneider's exoplanet
database {\tt www.exoplanet.eu}, and the Besan\c{c}on model of the
Galaxy at {\tt physique.obs-besancon.fr/modele/}. {\modif The
planetary evolution models used for this work can be downloaded at
www.obs-nice.fr/guillot/pegasids/}.

\bibliographystyle{aa}
\bibliography{corotlux}

\end{document}